\documentclass[iop]{emulateapj}

\newcommand{\NumTransits}{four}

\newcommand{\cOneGJKs}{0.00780}
\newcommand{\cOnePlusGJKs}{0.00026}
\newcommand{\cOneMinusGJKs}{0.00026}
\newcommand{\cTwoGJKs}{-0.064}
\newcommand{\cTwoPlusGJKs}{0.004}
\newcommand{\cTwoMinusGJKs}{0.003}

\newcommand{\TransitDepthPercentAbstractGJKs}{1.459}
\newcommand{\TransitDepthPercentAbstractMinusGJKs}{0.029}
\newcommand{\TransitDepthPercentAbstractPlusGJKs}{0.030}

\newcommand{\JDOffsetZeroGJKs}{5375.8500}
\newcommand{\JDOffsetPlusZeroGJKs}{0.0002}
\newcommand{\JDOffsetMinusZeroGJKs}{0.0001}

\newcommand{\cOneGJJ}{0.00542}
\newcommand{\cOnePlusGJJ}{0.00017}
\newcommand{\cOneMinusGJJ}{0.00020}
\newcommand{\cTwoGJJ}{-0.034}
\newcommand{\cTwoPlusGJJ}{0.002}
\newcommand{\cTwoMinusGJJ}{0.002}

\newcommand{\TransitDepthPercentAbstractGJJ}{1.334}
\newcommand{\TransitDepthPercentAbstractMinusGJJ}{0.021}
\newcommand{\TransitDepthPercentAbstractPlusGJJ}{0.020}

\newcommand{\JDOffsetZeroGJJ}{5375.8500}
\newcommand{\JDOffsetPlusZeroGJJ}{0.0001}
\newcommand{\JDOffsetMinusZeroGJJ}{0.0002}

\newcommand{\cOneGJCHFourOn}{0.00512}
\newcommand{\cOnePlusGJCHFourOn}{0.00035}
\newcommand{\cOneMinusGJCHFourOn}{0.00031}
\newcommand{\cTwoGJCHFourOn}{-0.010}
\newcommand{\cTwoPlusGJCHFourOn}{0.008}
\newcommand{\cTwoMinusGJCHFourOn}{0.009}

\newcommand{\TransitDepthPercentAbstractGJCHFourOn}{1.290}
\newcommand{\TransitDepthPercentAbstractMinusGJCHFourOn}{0.043}
\newcommand{\TransitDepthPercentAbstractPlusGJCHFourOn}{0.050}

\newcommand{\JDOffsetZeroGJCHFourOn}{5416.9402}
\newcommand{\JDOffsetPlusZeroGJCHFourOn}{0.0002}
\newcommand{\JDOffsetMinusZeroGJCHFourOn}{0.0004}

\newcommand{\cOneGJJII}{0.00624}
\newcommand{\cOnePlusGJJII}{0.00028}
\newcommand{\cOneMinusGJJII}{0.00026}
\newcommand{\cTwoGJJII}{-0.035}
\newcommand{\cTwoPlusGJJII}{0.007}
\newcommand{\cTwoMinusGJJII}{0.008}

\newcommand{\TransitDepthPercentAbstractGJJII}{1.302}
\newcommand{\TransitDepthPercentAbstractMinusGJJII}{0.040}
\newcommand{\TransitDepthPercentAbstractPlusGJJII}{0.044}

\newcommand{\JDOffsetZeroGJJII}{5416.9404}
\newcommand{\JDOffsetPlusZeroGJJII}{0.0001}
\newcommand{\JDOffsetMinusZeroGJJII}{0.0001}

\newcommand{\cOneGJKsIIQuadraticOLAM}{0.00703}
\newcommand{\cOnePlusGJKsIIQuadraticOLAM}{0.00026}
\newcommand{\cOneMinusGJKsIIQuadraticOLAM}{0.00026}
\newcommand{\cTwoGJKsIIQuadraticOLAM}{-0.098}
\newcommand{\cTwoPlusGJKsIIQuadraticOLAM}{0.007}
\newcommand{\cTwoMinusGJKsIIQuadraticOLAM}{0.002}
\newcommand{\cThreeGJKsIIQuadraticOLAM}{0.442}
\newcommand{\cThreePlusGJKsIIQuadraticOLAM}{0.029}
\newcommand{\cThreeMinusGJKsIIQuadraticOLAM}{0.056}

\newcommand{\TransitDepthPercentAbstractGJKsIIQuadraticOLAM}{1.422}
\newcommand{\TransitDepthPercentAbstractMinusGJKsIIQuadraticOLAM}{0.034}
\newcommand{\TransitDepthPercentAbstractPlusGJKsIIQuadraticOLAM}{0.032}

\newcommand{\JDOffsetZeroGJKsIIQuadraticOLAM}{5424.8423}
\newcommand{\JDOffsetPlusZeroGJKsIIQuadraticOLAM}{0.0001}
\newcommand{\JDOffsetMinusZeroGJKsIIQuadraticOLAM}{0.0001}

\newcommand{\cOneGJJIIIOLAM}{0.00739}
\newcommand{\cOnePlusGJJIIIOLAM}{0.00020}
\newcommand{\cOneMinusGJJIIIOLAM}{0.00023}
\newcommand{\cTwoGJJIIIOLAM}{-0.060}
\newcommand{\cTwoPlusGJJIIIOLAM}{0.002}
\newcommand{\cTwoMinusGJJIIIOLAM}{0.003}

\newcommand{\TransitDepthPercentAbstractGJJIIIOLAM}{1.368}
\newcommand{\TransitDepthPercentAbstractMinusGJJIIIOLAM}{0.021}
\newcommand{\TransitDepthPercentAbstractPlusGJJIIIOLAM}{0.026}

\newcommand{\JDOffsetZeroGJJIIIOLAM}{5424.8424}
\newcommand{\JDOffsetPlusZeroGJJIIIOLAM}{0.0004}
\newcommand{\JDOffsetMinusZeroGJJIIIOLAM}{0.0002}

\newcommand{\cOneGJKsLastOLAM}{0.00406}
\newcommand{\cOnePlusGJKsLastOLAM}{0.00032}
\newcommand{\cOneMinusGJKsLastOLAM}{0.00035}
\newcommand{\cTwoGJKsLastOLAM}{0.016}
\newcommand{\cTwoPlusGJKsLastOLAM}{0.006}
\newcommand{\cTwoMinusGJKsLastOLAM}{0.007}

\newcommand{\TransitDepthPercentAbstractGJKsLastOLAM}{1.424}
\newcommand{\TransitDepthPercentAbstractMinusGJKsLastOLAM}{0.031}
\newcommand{\TransitDepthPercentAbstractPlusGJKsLastOLAM}{0.044}

\newcommand{\JDOffsetZeroGJKsLastOLAM}{5462.7722}
\newcommand{\JDOffsetPlusZeroGJKsLastOLAM}{0.0002}
\newcommand{\JDOffsetMinusZeroGJKsLastOLAM}{0.0002}

\newcommand{\cOneGJJLastOLAM}{0.00493}
\newcommand{\cOnePlusGJJLastOLAM}{0.00030}
\newcommand{\cOneMinusGJJLastOLAM}{0.00031}
\newcommand{\cTwoGJJLastOLAM}{-0.013}
\newcommand{\cTwoPlusGJJLastOLAM}{0.005}
\newcommand{\cTwoMinusGJJLastOLAM}{0.005}

\newcommand{\TransitDepthPercentAbstractGJJLastOLAM}{1.307}
\newcommand{\TransitDepthPercentAbstractMinusGJJLastOLAM}{0.031}
\newcommand{\TransitDepthPercentAbstractPlusGJJLastOLAM}{0.034}

\newcommand{\JDOffsetZeroGJJLastOLAM}{5462.7722}
\newcommand{\JDOffsetPlusZeroGJJLastOLAM}{0.0002}
\newcommand{\JDOffsetMinusZeroGJJLastOLAM}{0.0002}

\newcommand{\cOneGJJointAll}{0.00543}
\newcommand{\cOnePlusGJJointAll}{0.00019}
\newcommand{\cOneMinusGJJointAll}{0.00017}
\newcommand{\cTwoGJJointAll}{-0.034}
\newcommand{\cTwoPlusGJJointAll}{0.002}
\newcommand{\cTwoMinusGJJointAll}{0.002}

\newcommand{\TransitDepthPercentAbstractGJJointAll}{1.350}
\newcommand{\TransitDepthPercentAbstractMinusGJJointAll}{0.018}
\newcommand{\TransitDepthPercentAbstractPlusGJJointAll}{0.022}

\newcommand{\JDOffsetZeroGJJointAll}{5375.8501}
\newcommand{\JDOffsetPlusZeroGJJointAll}{0.0001}
\newcommand{\JDOffsetMinusZeroGJJointAll}{0.0001}

\newcommand{\ParamEightGJJointAll}{1.22}
\newcommand{\ParamEightPlusGJJointAll}{0.05}
\newcommand{\ParamEightMinusGJJointAll}{0.04}

\newcommand{\ParamTenGJJointAll}{-0.69}
\newcommand{\ParamTenPlusGJJointAll}{0.04}
\newcommand{\ParamTenMinusGJJointAll}{0.04}

\newcommand{\ParamTwelveGJJointAll}{1.07}
\newcommand{\ParamTwelvePlusGJJointAll}{0.03}
\newcommand{\ParamTwelveMinusGJJointAll}{0.05}

\newcommand{\ParamFourteenGJJointAll}{-0.57}
\newcommand{\ParamFourteenPlusGJJointAll}{0.04}
\newcommand{\ParamFourteenMinusGJJointAll}{0.03}
\newcommand{\ParamFifteenGJJointAll}{1.063}
\newcommand{\ParamFifteenPlusGJJointAll}{0.019}
\newcommand{\ParamFifteenMinusGJJointAll}{0.021}
\newcommand{\ParamSixteenGJJointAll}{0.00777}
\newcommand{\ParamSixteenPlusGJJointAll}{0.00024}
\newcommand{\ParamSixteenMinusGJJointAll}{0.00028}
\newcommand{\ParamSeventeenGJJointAll}{-0.065}
\newcommand{\ParamSeventeenPlusGJJointAll}{0.003}
\newcommand{\ParamSeventeenMinusGJJointAll}{0.003}
\newcommand{\JDOffsetEighteenGJJointAll}{5424.8423}
\newcommand{\JDOffsetPlusEighteenGJJointAll}{0.0001}
\newcommand{\JDOffsetMinusEighteenGJJointAll}{0.0001}
\newcommand{\ParamNineteenGJJointAll}{0.00708}
\newcommand{\ParamNineteenPlusGJJointAll}{0.00024}
\newcommand{\ParamNineteenMinusGJJointAll}{0.00029}
\newcommand{\ParamTwentyGJJointAll}{-0.097}
\newcommand{\ParamTwentyPlusGJJointAll}{0.010}
\newcommand{\ParamTwentyMinusGJJointAll}{0.003}
\newcommand{\ParamTwentyOneGJJointAll}{0.00734}
\newcommand{\ParamTwentyOnePlusGJJointAll}{0.00020}
\newcommand{\ParamTwentyOneMinusGJJointAll}{0.00021}
\newcommand{\ParamTwentyTwoGJJointAll}{-0.060}
\newcommand{\ParamTwentyTwoPlusGJJointAll}{0.002}
\newcommand{\ParamTwentyTwoMinusGJJointAll}{0.002}
\newcommand{\JDOffsetTwentyThreeGJJointAll}{5462.7722}
\newcommand{\JDOffsetPlusTwentyThreeGJJointAll}{0.0001}
\newcommand{\JDOffsetMinusTwentyThreeGJJointAll}{0.0001}
\newcommand{\ParamTwentyFourGJJointAll}{0.00406}
\newcommand{\ParamTwentyFourPlusGJJointAll}{0.00028}
\newcommand{\ParamTwentyFourMinusGJJointAll}{0.00038}
\newcommand{\ParamTwentyFiveGJJointAll}{0.015}
\newcommand{\ParamTwentyFivePlusGJJointAll}{0.006}
\newcommand{\ParamTwentyFiveMinusGJJointAll}{0.006}
\newcommand{\ParamTwentySixGJJointAll}{0.00497}
\newcommand{\ParamTwentySixPlusGJJointAll}{0.00027}
\newcommand{\ParamTwentySixMinusGJJointAll}{0.00031}
\newcommand{\ParamTwentySevenGJJointAll}{-0.012}
\newcommand{\ParamTwentySevenPlusGJJointAll}{0.005}
\newcommand{\ParamTwentySevenMinusGJJointAll}{0.006}
\newcommand{\ParamTwentyEightGJJointAll}{1.364}
\newcommand{\ParamTwentyEightPlusGJJointAll}{0.021}
\newcommand{\ParamTwentyEightMinusGJJointAll}{0.024}
\newcommand{\ParamTwentyNineGJJointAll}{1.329}
\newcommand{\ParamTwentyNinePlusGJJointAll}{0.026}
\newcommand{\ParamTwentyNineMinusGJJointAll}{0.028}
\newcommand{\ParamThirtyGJJointAll}{0.426}
\newcommand{\ParamThirtyPlusGJJointAll}{0.037}
\newcommand{\ParamThirtyMinusGJJointAll}{0.066}
\newcommand{\DiffSigmaIntJointKsJGJJointAll}{3}
\newcommand{\TransitDepthPercentAbstractKsTwoEightGJJointAll}{1.450}
\newcommand{\TransitDepthPercentAbstractKsTwoEightMinusGJJointAll}{0.036}
\newcommand{\TransitDepthPercentAbstractKsTwoEightPlusGJJointAll}{0.036}
\newcommand{\TransitDepthPercentAbstractKsTwoNineGJJointAll}{1.412}
\newcommand{\TransitDepthPercentAbstractKsTwoNineMinusGJJointAll}{0.039}
\newcommand{\TransitDepthPercentAbstractKsTwoNinePlusGJJointAll}{0.039}
\newcommand{\TransitDepthPercentAbstractKsSixGJJointAll}{1.435}
\newcommand{\TransitDepthPercentAbstractKsSixMinusGJJointAll}{0.034}
\newcommand{\TransitDepthPercentAbstractKsSixPlusGJJointAll}{0.034}

\newcommand{\LimbJCTwo}{1.24}
\newcommand{\LimbJCFour}{-0.67}
\newcommand{\LimbKsCTwo}{1.06}
\newcommand{\LimbKsCFour}{-0.59}
\newcommand{\LimbCHFourOnCTwo}{1.22}
\newcommand{\LimbCHFourOnCFour}{-0.66}

\newcommand{\GJJBandCombinedTransitDepthPercent}{1.338}
\newcommand{\GJJBandCombinedTransitDepthPercentMinusPlus}{0.013}
\newcommand{\GJKsBandCombinedTransitDepthPercent}{1.438}
\newcommand{\GJKsBandCombinedTransitDepthPercentMinusPlus}{0.019}
\newcommand{\GJKsJSigmaDiff}{5}

\newcommand{\ZetaJ}{1.02}

\newcommand{\ChiHTwoO}{26}
\newcommand{\ChiSolar}{9}

\newcommand{\ChiNoMethane}{7}
\newcommand{\NSigmaModels}{2}

\newcommand{\ChiHTwoOEVERYTHING}{40}
\newcommand{\ChiSolarEVERYTHING}{101}

\newcommand{\ChiHazeEVERYTHING}{34}

\newcommand{\GJKsOverJCombined}{1.072}
\newcommand{\GJKsOverJCombinedMinusPlus}{0.018}

\newcommand{\GJKsOverJSigmaDiff}{4}

\newcommand{\GJKsOverJChiSquared}{1.92}
\newcommand{\RadiusKsIncreaseGJKsOverJ}{1.04}

\newcommand{\KilometersRadiusKsIncreaseRoundGJKsOverJ}{610}

\shorttitle{Broadband Transmission Spectrum of GJ 1214b} 
\shortauthors{Croll et al.}

\begin{document}
\title{Broadband Transmission Spectroscopy of the super-Earth GJ 1214b suggests a low mean molecular weight atmosphere\altaffilmark{*}}


\author{Bryce Croll\altaffilmark{1},
Loic Albert\altaffilmark{2},
Ray Jayawardhana\altaffilmark{1},
Eliza Miller-Ricci Kempton\altaffilmark{3},
Jonathan J. Fortney\altaffilmark{3},
Norman Murray\altaffilmark{4,5},
Hilding Neilson\altaffilmark{6}
}

\altaffiltext{1}{Department of Astronomy and Astrophysics, University of Toronto, 50 St. George Street, Toronto, ON 
M5S 3H4, Canada;
croll@astro.utoronto.ca}

\altaffiltext{2}{D\'epartement de physique, Universit\'e de Montr\'eal, C.P.
6128 Succ. Centre-Ville, Montr\'eal, QC, H3C 3J7, Canada}

\altaffiltext{3}{Department of Astronomy and Astrophysics, University of California, Santa Cruz, CA, 95064}

\altaffiltext{4}{Canadian Institute for Theoretical Astrophysics, 60 St. George Street, University of Toronto, Toronto ON M5S 3H8, Canada}

\altaffiltext{5}{Canada Research Chair in Astrophysics}

\altaffiltext{6}{Argelander-Institut f\"{u}r Astronomie, Auf dem H\"{u}gel 71, D-53121 Bonn, Germany}

\altaffiltext{*}{Based on observations obtained with WIRCam, a joint project of CFHT, Taiwan, Korea, Canada, France, at the Canada-France-Hawaii Telescope (CFHT) which is operated by the National Research Council (NRC) of Canada, the Institute National des Sciences de l'Univers of the Centre National de la Recherche Scientifique of France, and the University of Hawaii.}


\begin{abstract}

We used the Wide-field Infrared Camera on the Canada-France-Hawaii telescope
to observe \NumTransits \ transits of the super-Earth planet GJ 1214b in the near-infrared.
For each transit we observed in two bands nearly-simultaneously by rapidly switching the WIRCam filter wheel
back and forth
for the duration of the observations.
By combining all our J-band ($\sim$1.25 $\mu m$) observations we find a transit depth, analogous to the planet-to-star radius ratio squared,
in this band of $(R_{PJ}/R_{*})^2$=\GJJBandCombinedTransitDepthPercent $\pm$\GJJBandCombinedTransitDepthPercentMinusPlus\%
-- a value consistent with the optical transit depth reported by Charbonneau and collaborators.
However, our best-fit combined Ks-band ($\sim$2.15 $\mu m$) transit depth is deeper:
$(R_{PKs}/R_{*})^2$=\GJKsBandCombinedTransitDepthPercent $\pm$\GJKsBandCombinedTransitDepthPercentMinusPlus\%.
Formally our Ks-band transits are deeper than the J-band transits observed simultaneously
by a factor of 
$(R_{PKs}/R_{PJ})^2$=\GJKsOverJCombined $\pm$\GJKsOverJCombinedMinusPlus \
- a \GJKsOverJSigmaDiff$\sigma$  discrepancy.
The most straightforward explanation for our deeper Ks-band transit depth is a spectral absorption feature from the limb of the
atmosphere of the planet;
for the spectral absorption feature to be this prominent 
the atmosphere of GJ 1214b must have a large scale height and a low mean molecular weight.
That is, its atmosphere would have to be hydrogen/helium dominated and this planet would be better described as a mini-Neptune. 
However, recently published observations from 0.78 - 1.0 $\mu m$, by Bean and collaborators,
show a lack of spectral features and transit depths consistent with those obtained
by Charbonneau and collaborators. The most likely atmospheric composition for GJ 1214b
that arises from combining
all these observations is less clear; if the atmosphere of GJ 1214b is hydrogen/helium dominated then it must have
either a haze layer that is obscuring transit depth differences at shorter wavelengths,
or significantly different spectral features than current models predict.
Our observations disfavour a water-world composition, but such a composition 
will remain a possibility for GJ 1214b,
until observations reconfirm our deeper Ks-band transit depth or detect features at other wavelengths.
\end{abstract}


\section{Introduction}

	Astronomers have been waiting for sometime for a planet remotely similar to our own Earth
that could be readily investigated with current instruments.
Such an object was recently announced with the 
seminal discovery of the super-Earth planet 
GJ 1214b \citep{Charbonneau09} with the MEarth telescope network \citep{NutzmanCharbonneau08,Irwin09}.
Although not the first transiting super-Earth announced -- CoRoT-7b arguably holds that honour 
\citep{Leger09,Queloz09,Pont10} -- 
GJ 1214b is in many ways more interesting as it offers the opportunity for advantageous
follow-up to constrain its planetary characteristics.
With a mass
of 6.55 $M_{\earth}$ and a radius of 2.68 $R_{\earth}$, GJ 1214b's density ($\rho$$\sim$1.87 g $cm^{-3}$; \citealt{Charbonneau09})
is less than that of the
terrestrial planets of our solar system and therefore GJ 1214b may 
have a significant gaseous atmosphere.
Also, as it transits a low mass star,
its equilibrium temperature ($T_{eq}$$\sim$500K assuming a low Bond albedo) is much more hospitable
than CoRoT-7b, and it has a much 
more favourable planet-to-star radius ratio; as a result,
if there are significant spectral features in its atmosphere
then they should be detectable with current instruments.

	As there are no super-Earth analogues in our solar system, it is a pressing question 
whether the burgeoning class of planets with minimum
masses below 10$M_{\earth}$  (e.g. \citealt{Udry07,Mayor09a,Mayor09b}) are predominantly 
scaled-down Neptunes, with large helium-hydrogen atmospheric
envelopes, or scaled-up terrestrial planets with atmospheres predominately composed of heavier molecules.
Fortunately, GJ 1214b
is an ideal candidate to answer such questions; 
\citet{MillerRicciFortney10} showed that due to GJ 1214b's advantageous scale-height
and planet-to-star radius ratio it should have 
readily observable water and methane spectral features across the infrared spectrum
if its atmosphere is composed primarily of hydrogen and helium.
Conversely,
if its atmosphere is composed predominantly of heavier molecules, then the resulting
smaller scale height will mute the 
spectral features and current instruments will return transit depths consistent 
with the depths measured in the optical by 
\citet{Charbonneau09}.

Furthermore, recent theoretical work has shown
that measurements to constrain the composition of the gaseous atmosphere of GJ 1214b will also
constrain the planet's bulk composition.
\citet{RogersSeager10} and later \citet{Nettelmann10} showed that 
the observed mass and radius \citep{Charbonneau09}
can be equally well fit by either a significant rocky core/mantle
and a 
hydrogen-rich atmosphere, or a ``water-world'' with a 
small water-rich core and a significant steam atmosphere. 
Searches for spectral features will differentiate between these two scenarios; 
detections of prominent spectral features
will argue for the former scenario of a rocky core to go along with the hydrogen/helium-rich
gaseous planetary envelope.

	Recently, \citet{Bean10} have performed just such transit spectroscopy observations using
the FORS2 instrument on the VLT; \citet{Bean10} obtained 11 spectrophotometric light curves from 0.78 - 1.0 
$\mu m$ that show consistent transit depths with one another. By comparing to the \citet{MillerRicciFortney10} atmospheric models 
they were able to show that the lack of observed spectral features suggest that GJ 1214b must either have
a high mean molecular weight and is likely a water-world, or its atmosphere is hydrogen/helium dominated with 
hazes or clouds high in the atmosphere that obscure the expected spectral features shortward of $\sim$1 micron.
An additional possibility could certainly be that  
GJ 1214b's atmosphere is 
more complicated than expected, and its atmosphere 
could still be hydrogen/helium dominated
with different spectral features than 
the \citet{MillerRicciFortney10}
models suggest. 

	We have also performed broadband transmission spectroscopy observations
searching for GJ 1214b's spectral features from $\sim$1-2.5 microns
using the
Wide-field Infrared Camera (WIRCam) on the Canada-France-Hawaii telescope (CFHT). 
We have already successfully demonstrated the precision of WIRCam on CFHT in the near-infrared
through our detections of the secondary eclipses and 
thermal emission for TrES-2b and TrES-3b in the Ks-band \citep{CrollTrESTwo,CrollTrESThree},
and for WASP-12b in the J, H \& Ks-bands \citep{CrollWASPTwelve}. Here we report 
observations of several transits of the super-Earth GJ 1214b
in three bands with WIRCam on CFHT; for each transit we observed near-simultaneously in two bands
to allow for accurate comparisons of the transit depths between these two bands.
We observe an increased transit depth in the Ks-band
as compared to the
J-band depth, likely indicative of absorption near $\sim$2.15 $\mu$m.
The only way to achieve an absorption feature this prominent is if
GJ 1214b has a large scale height, a 
low mean molecular weight, and thus its atmosphere is hydrogen/helium dominated.
We discuss below the likely possibilities 
for the atmospheric make-up of GJ 1214b that result from a combination of the \citet{Bean10}, \citet{Charbonneau09}
and our own data. 
Our results disfavour a water-world composition, but such a composition is possible if our Ks-band point is simply an outlier;
such a composition will remain a distinct possibility
until further observations either confirm our increased Ks-band depth or detect spectral features at other wavelengths.
The observations to date
are most qualitatively consistent with a hydrogen/helium
dominated atmosphere that is either hazy or one with more complicated spectral features
than our current models suggest, such as an atmosphere where non-equilibrium chemistry plays a significant role.

\section{Observations and Data Reduction}
\label{SecReduction}

\begin{figure*}
\centering
\includegraphics[scale=0.44, angle = 270]{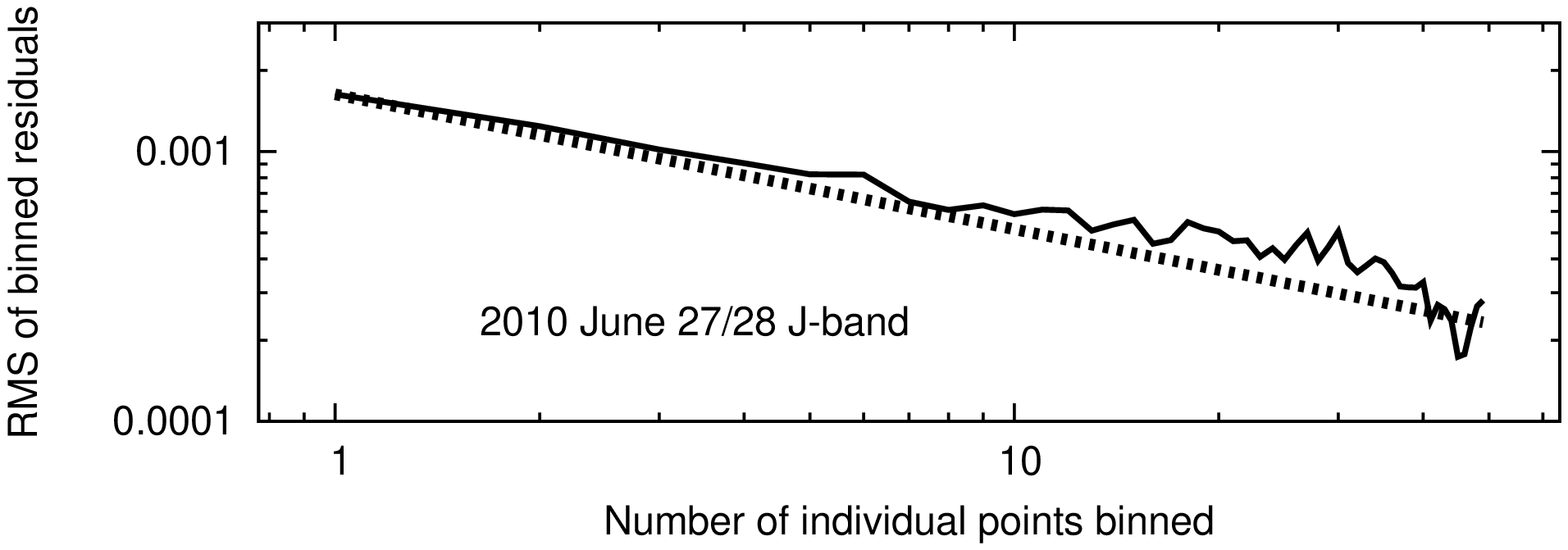}
\includegraphics[scale=0.44, angle = 270]{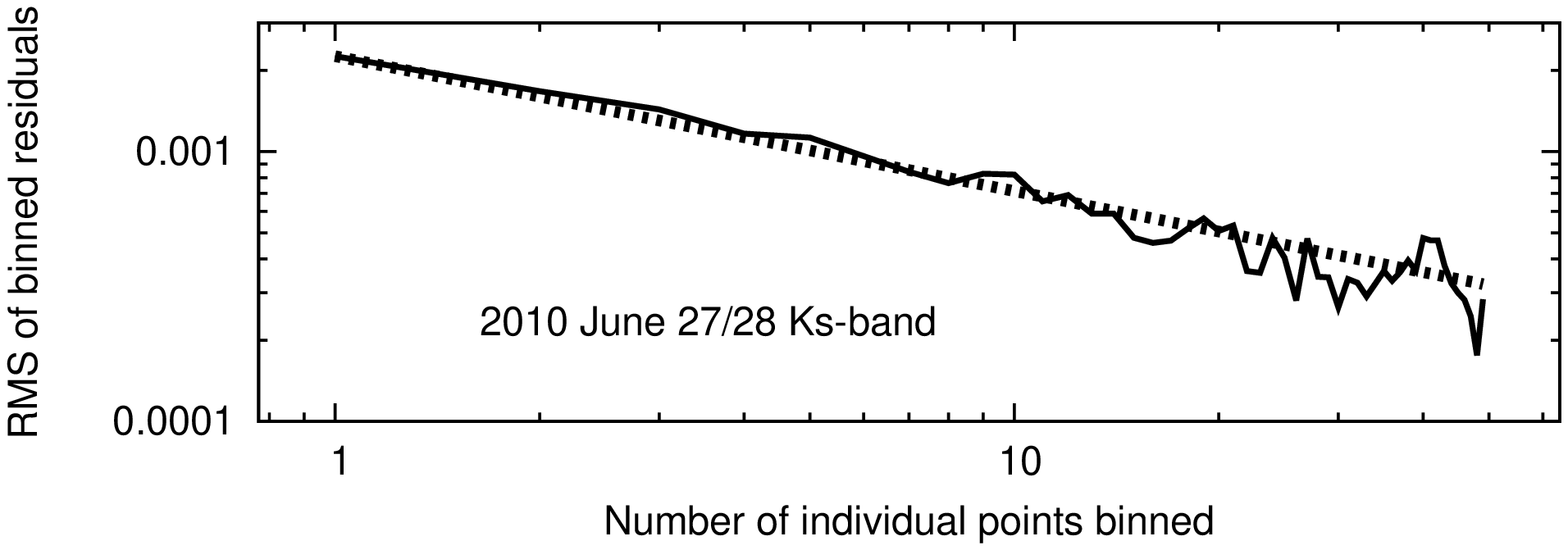}
\includegraphics[scale=0.44, angle = 270]{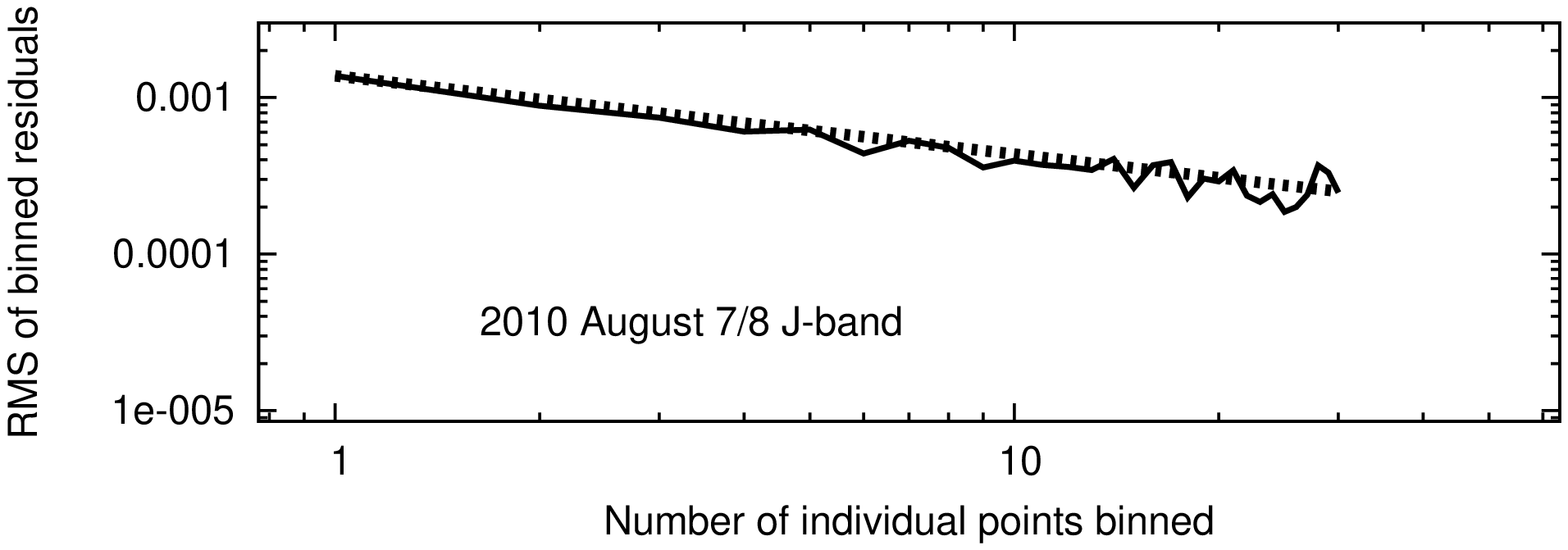}
\includegraphics[scale=0.44, angle = 270]{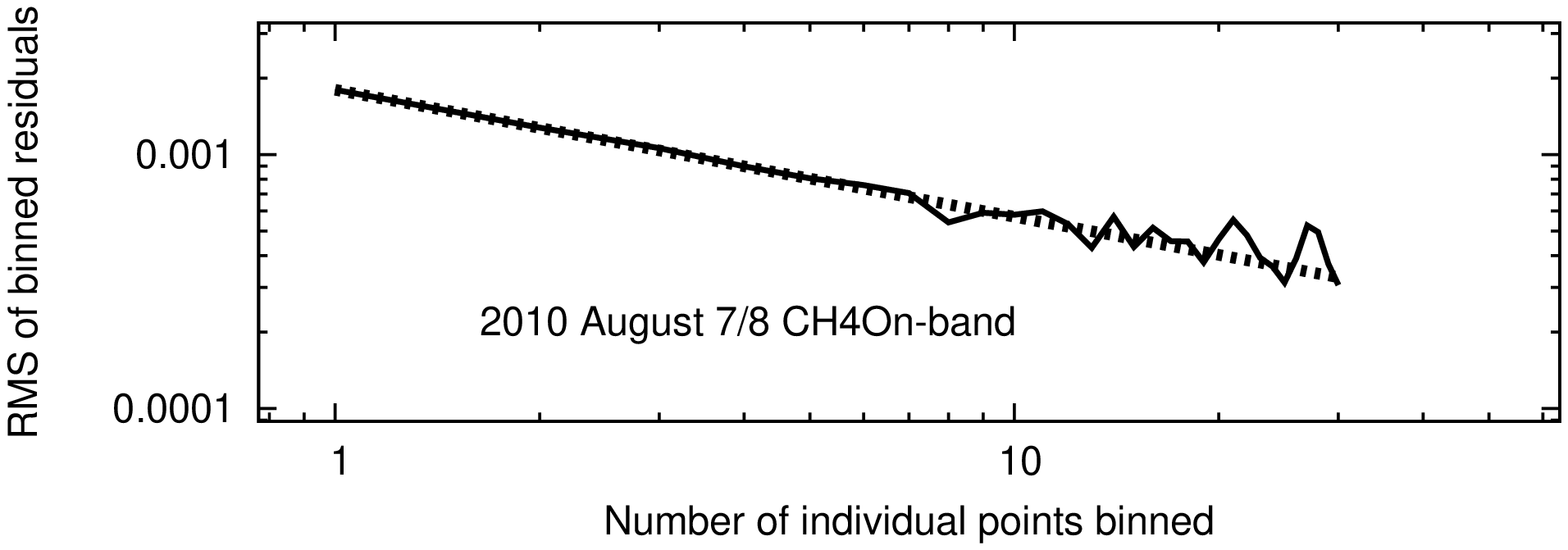}
\includegraphics[scale=0.44, angle = 270]{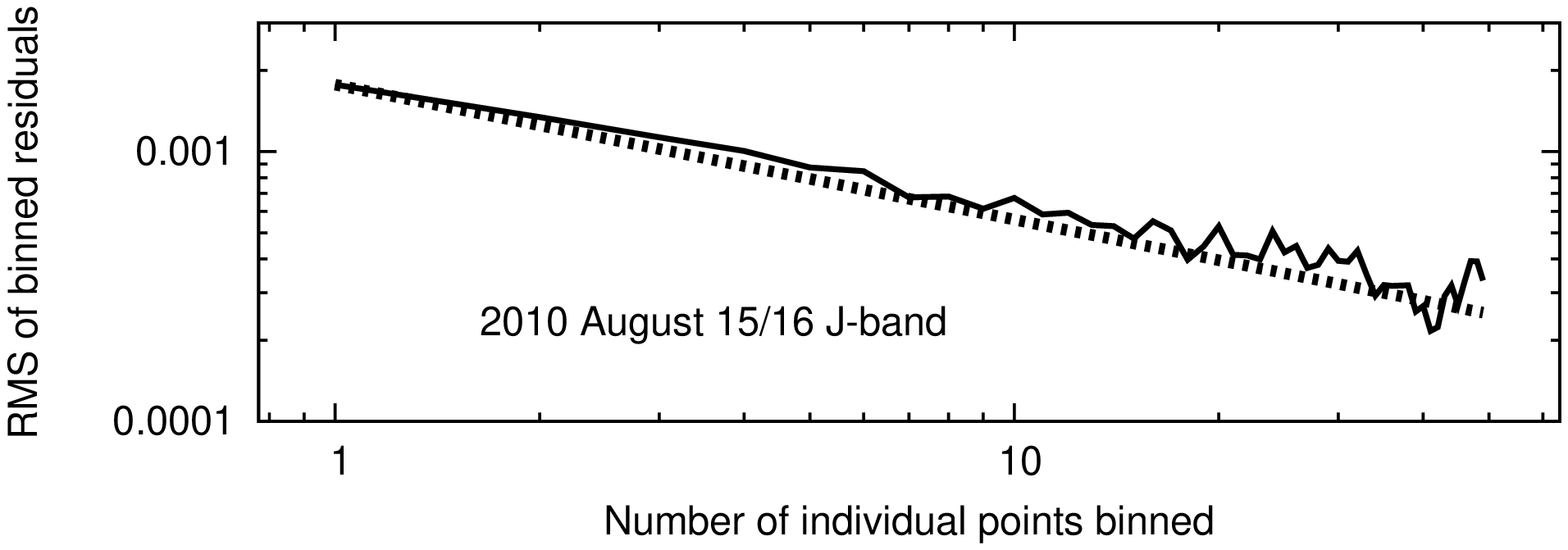}
\includegraphics[scale=0.44, angle = 270]{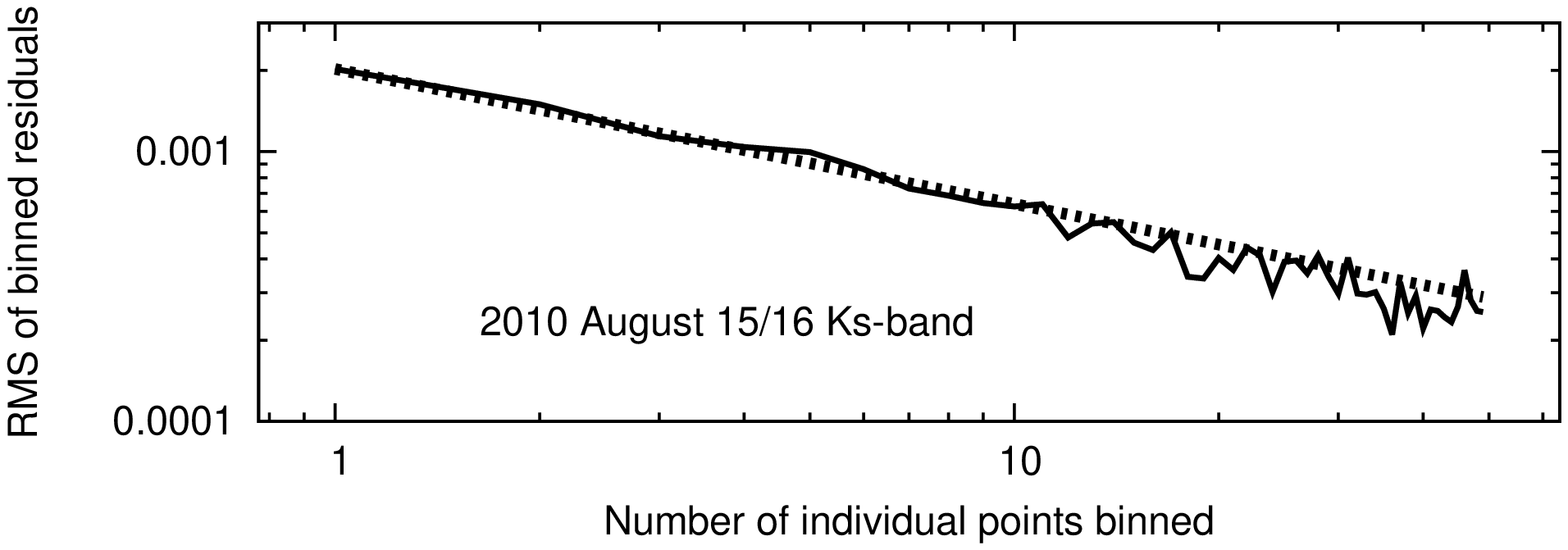}
\includegraphics[scale=0.44, angle = 270]{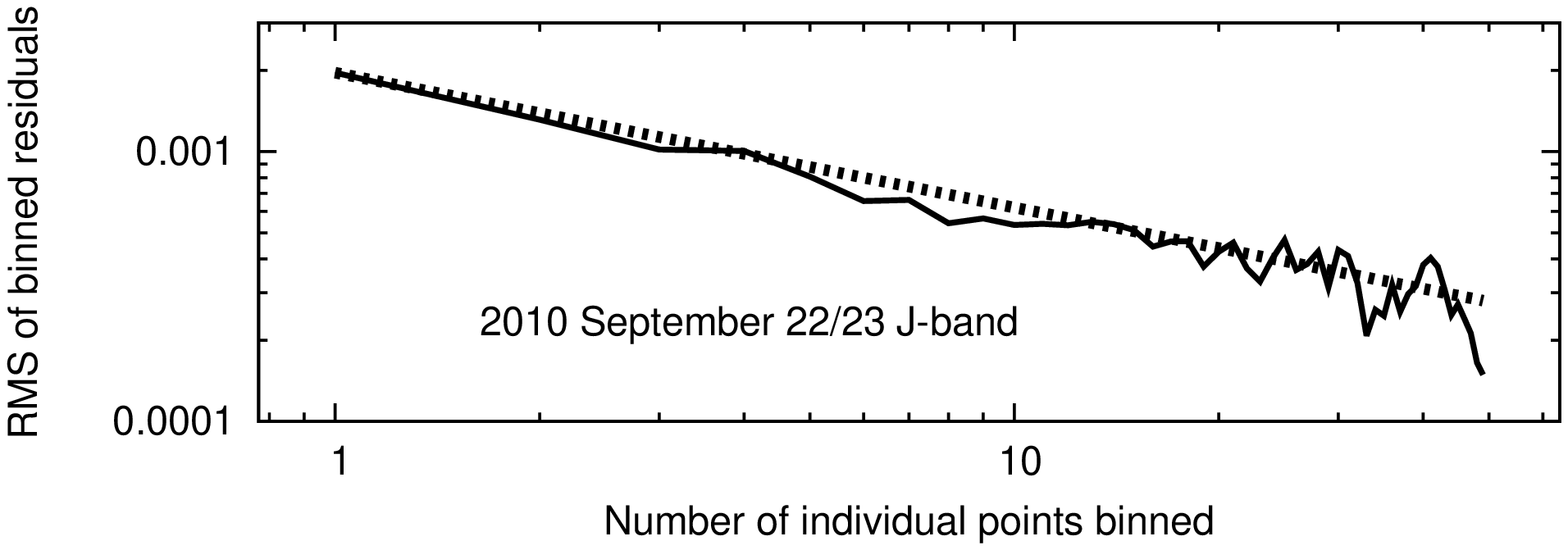}
\includegraphics[scale=0.44, angle = 270]{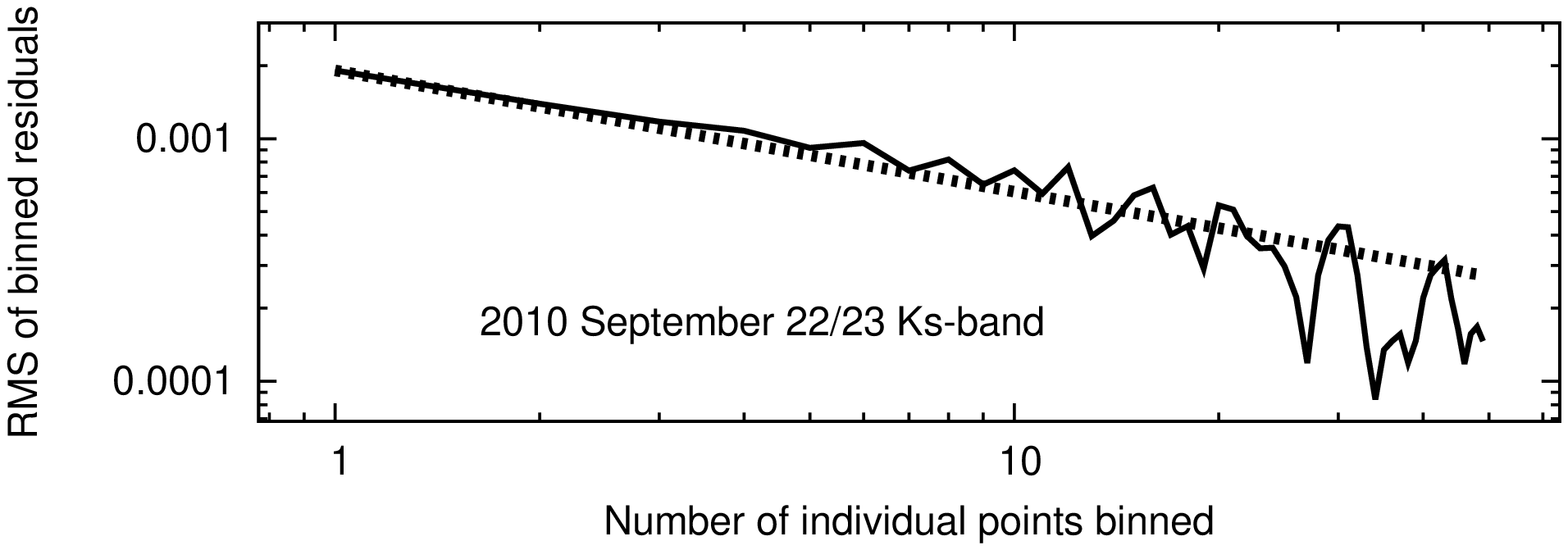}

\caption{	The root-mean-square of our out-of-eclipse photometry (solid line) 
		after the subtraction of their respective background trends
		for our various data-sets.
		The dashed line in each panel displays the
		one over the square-root of the bin-size expectation for Gaussian noise.
		}
\label{FigPoisson}
\end{figure*}

%

\begin{figure*}
\centering
\includegraphics[scale=0.42,angle=270]{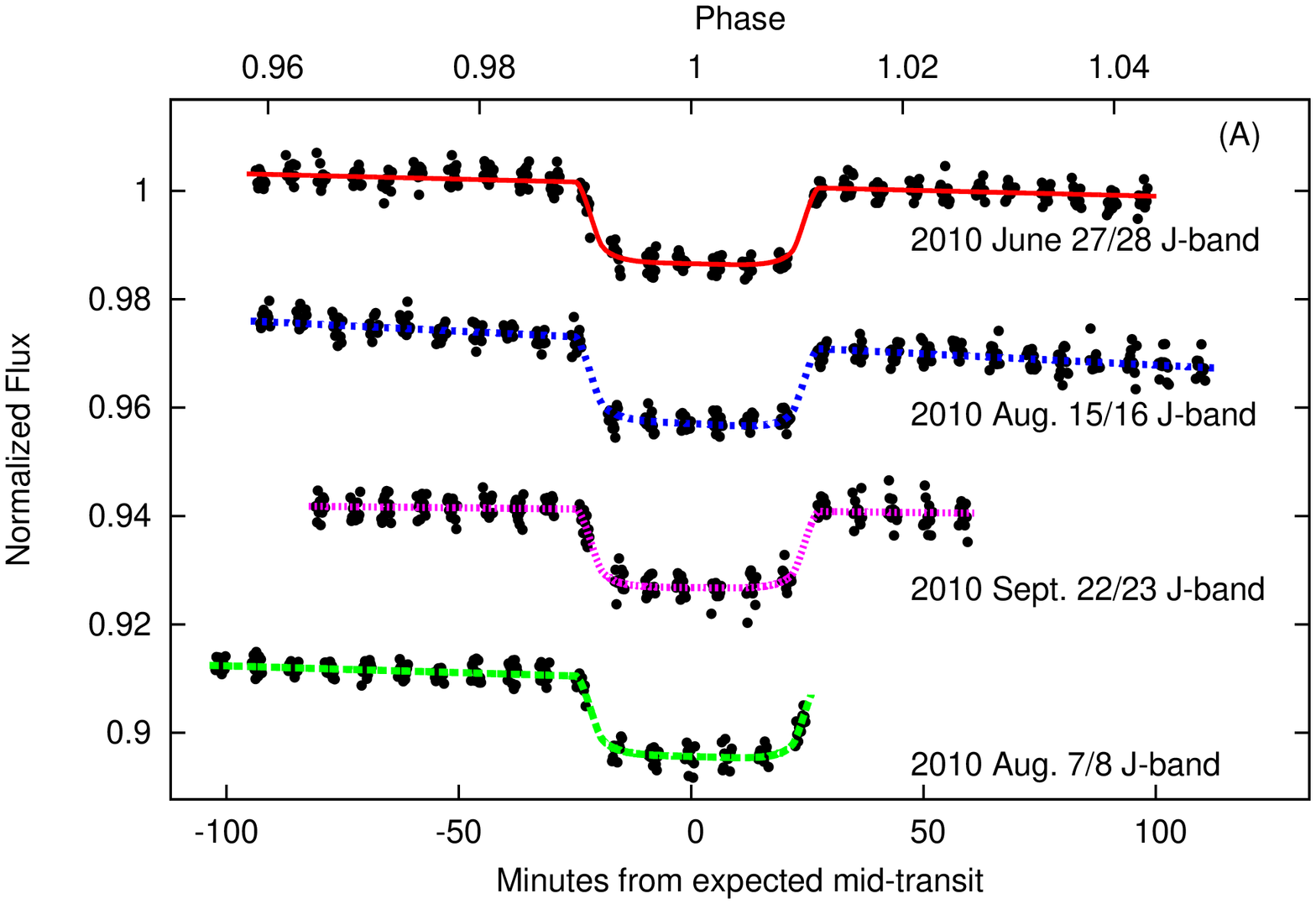}
\includegraphics[scale=0.42,angle=270]{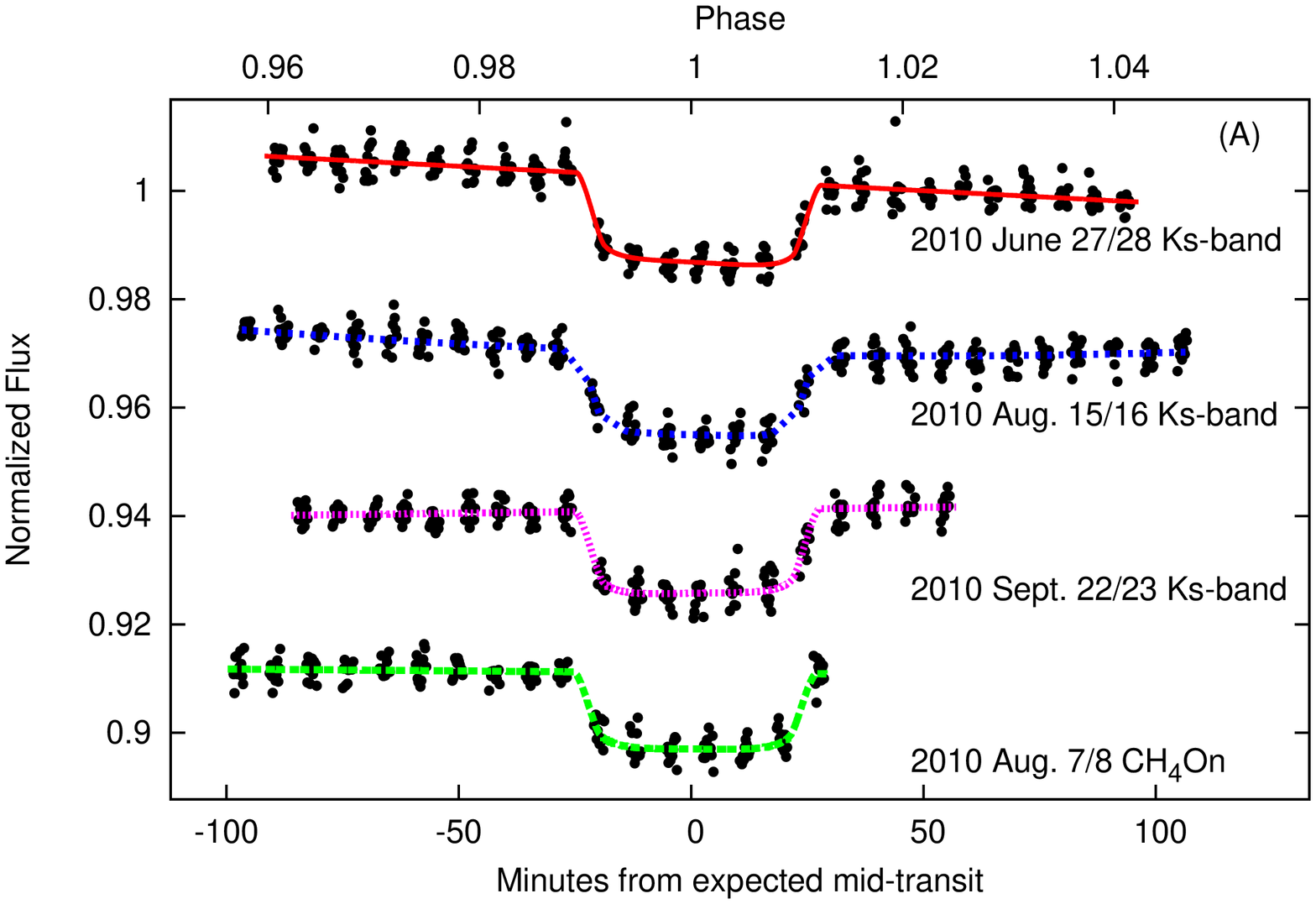}
\includegraphics[scale=0.42,angle=270]{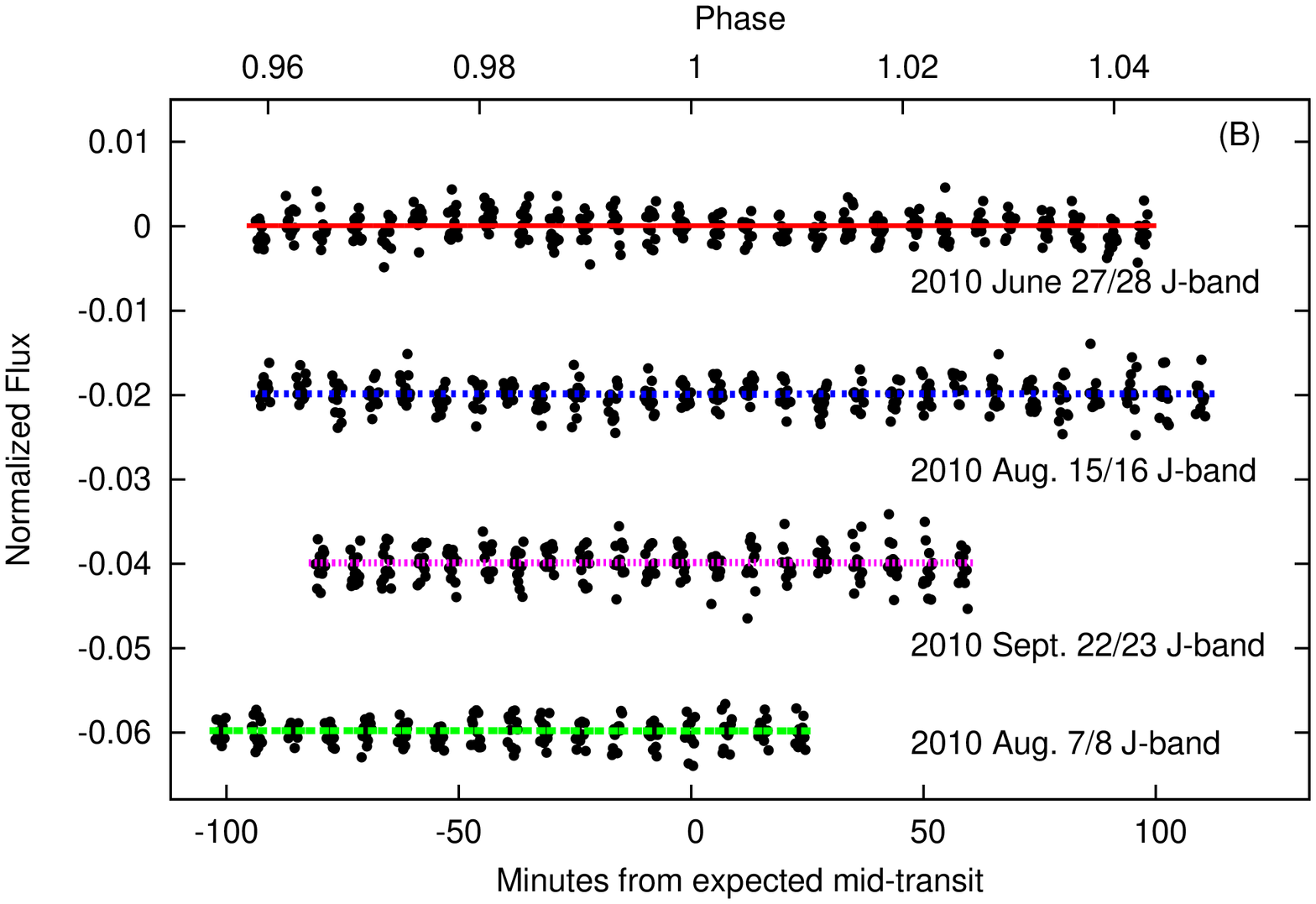}
\includegraphics[scale=0.42,angle=270]{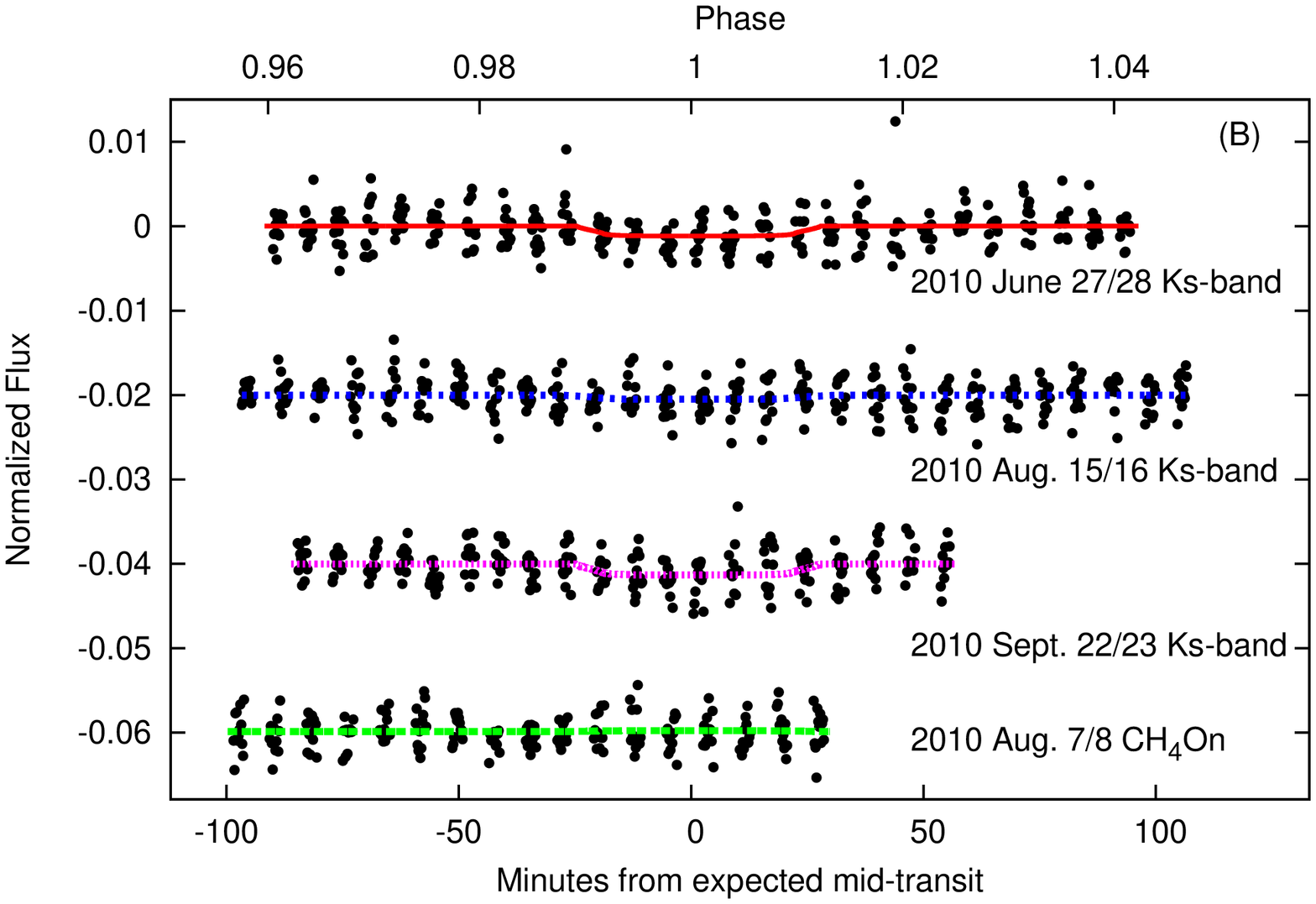}
\includegraphics[scale=0.42,angle=270]{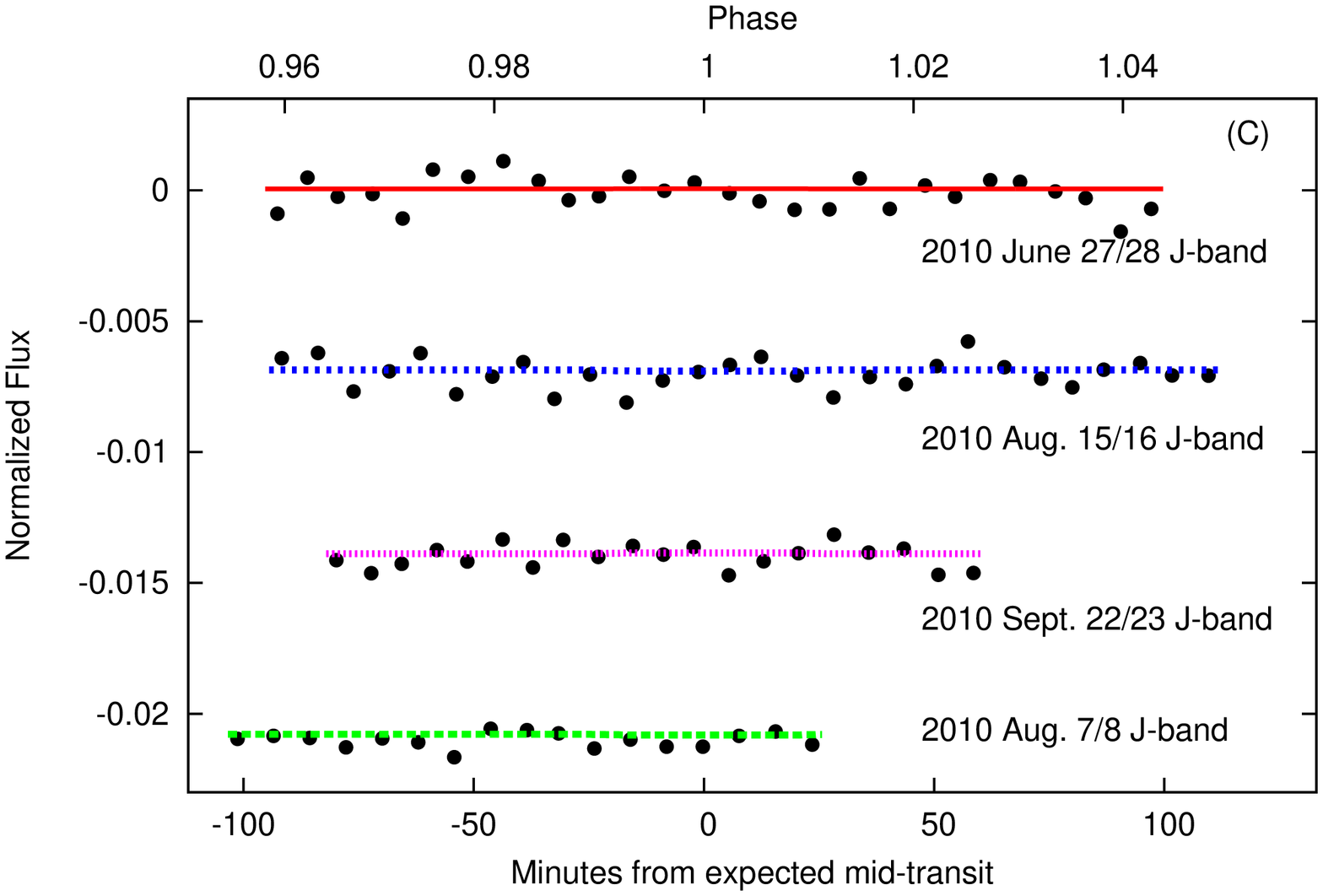}
\includegraphics[scale=0.42,angle=270]{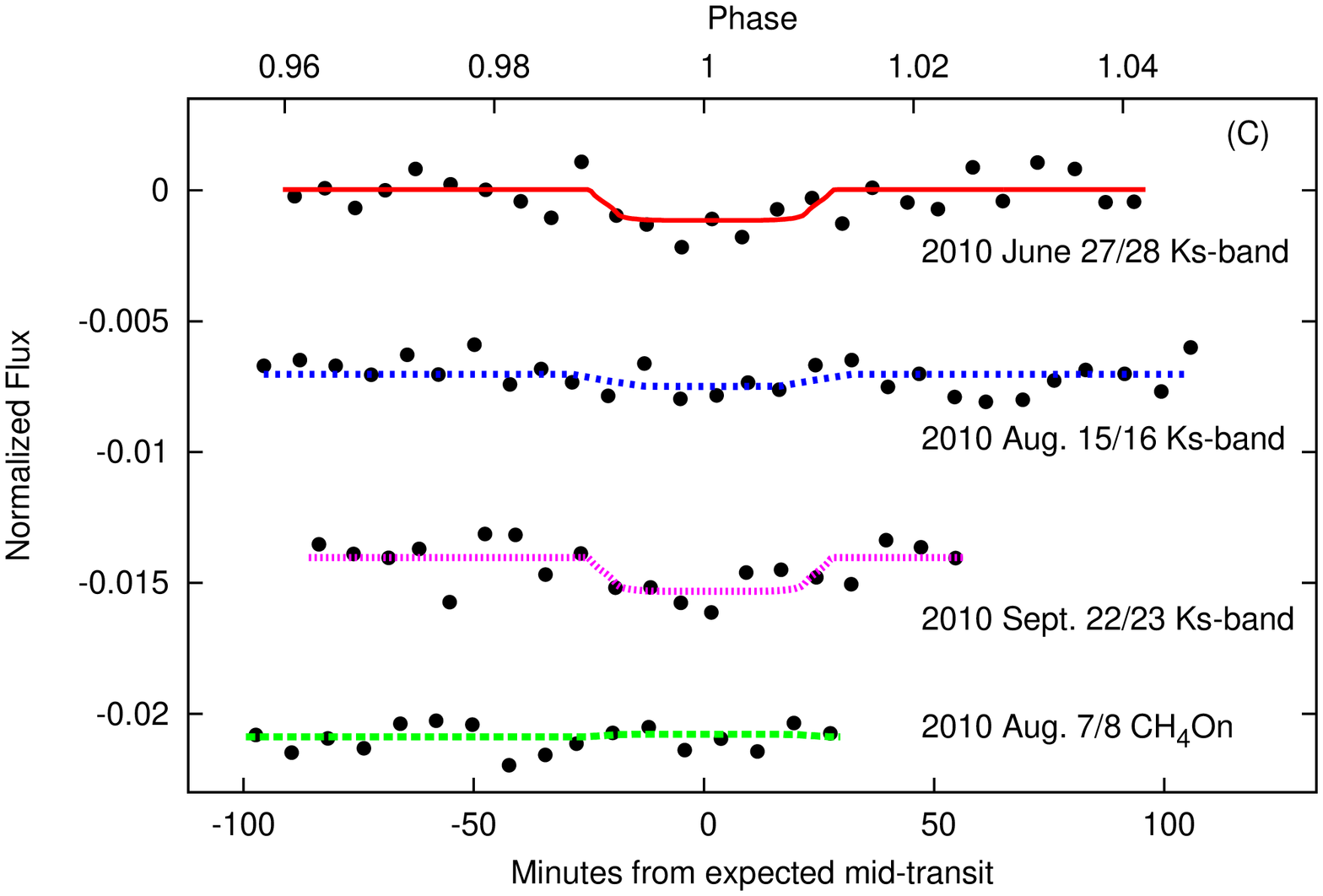}
\includegraphics[scale=0.40,angle=270]{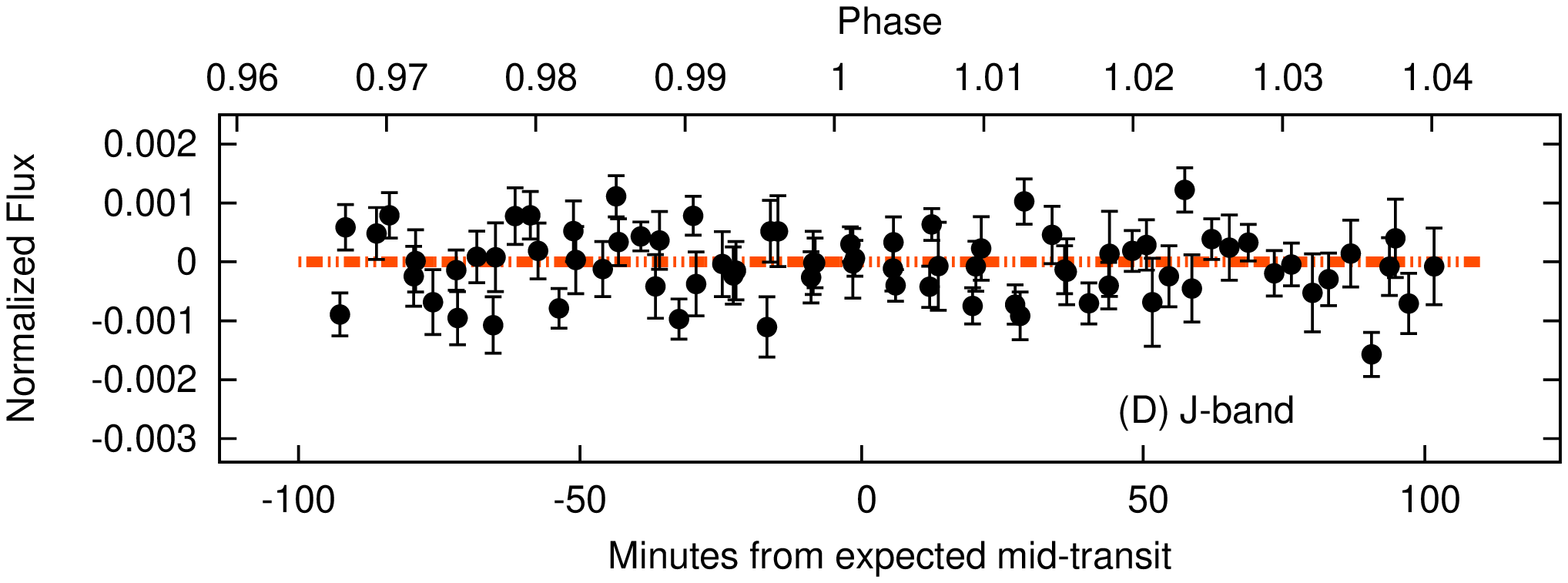}
\includegraphics[scale=0.40,angle=270]{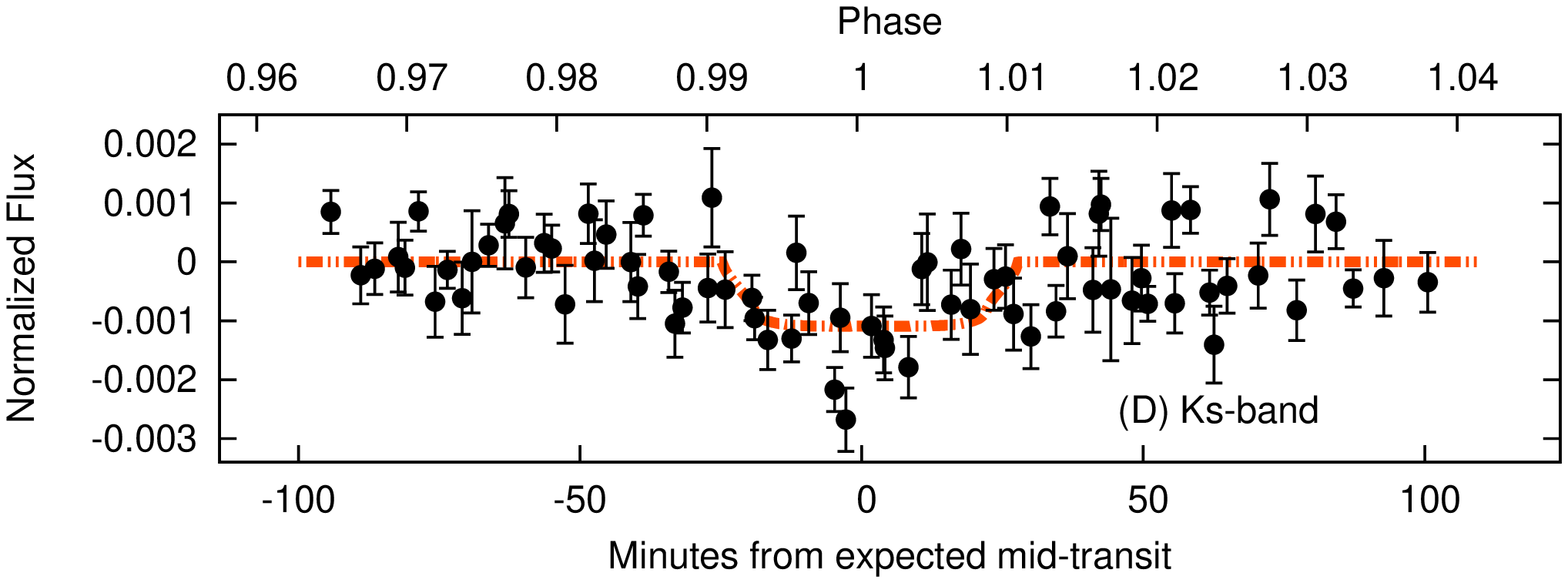}
\caption{ 	CFHT/WIRCam photometry bracketing the transit of GJ 1214
		obtained nearly simultaneously on the nights denoted in the figure
		in the J-band (left panels) 
		and the Ks-band and the CH$_4$On filter (right panels).
		The top set of panels (A) show the data and the best-fit transit
		model displayed with the appropriately coloured line.
		The second set of panels (B) show the residuals
		of the data and the models from a transit model with a depth 
		given by the best-fit value of the individual J-bands transits
		at left.
		Thus the (B) left panel simply displays the residuals from the best-fit transit model
		from that night's data.
		The (B) right panel displays the residuals of the Ks and CH$_4$On
		filter data and model 
		from a transit with a depth equal to the best-fit J-band transit depth observed
		on that night nearly simultaneously and using the appropriate Ks or CH$_4$On limb-darkening coefficients.
		The third set of panels (C) are the same as the (B) panels, except with the data
		binned every twelve points. 
		The bottom panels (D) display all the data obtained simultaneously in the J and Ks-bands from the (C) panels
		binned every twelve points [that is the top three transit curves in the ABC panels];
		the orange dot-dash line displays the difference from the best-fit J-band transit
		model using the combined J and Ks-band best-fit depths ($\S$\ref{SecDiscuss}).
		The errors in this panel are calculated from
		the standard deviation of the points within each bin.
		The best-fit Ks-band transits display increased transit depths, while the best-fit CH$_4$On
		transit displays a smaller transit depth than the J-band transits observed simultaneously.
		In the top three panels the different data-sets are offset vertically for clarity.
		Note each set of panels has a different vertical scale.
	}
\label{FigTransits}
\end{figure*}

 We observed the transit of GJ 1214b on \NumTransits \ occasions with WIRCam on CFHT.
For each transit we observed GJ 1214 in two bands
nearly simultaneously by rapidly switching the WIRCam filter wheel
back and forth between the two filters.
We observed GJ 1214 alternating between the J ($\sim$1.25 $\mu m$) and Ks ($\sim$2.15 $\mu m$) filters
on the evenings of 2010 June 27, 2010 August 15 and 2010 September 22.
On the evening of 2010 August 7 we observed GJ 1214 alternating between the 
J and CH$_4$On filters ($\sim$1.69 $\mu m$);
these 2010 August 7 observations 
were of a partial transit only,
as the airmass of GJ 1214 increased during transit
and during egress reached an airmass of $\sim$2.3, close
to the hard limit of the telescope, preventing further observations.
For the 2010 June 27 transit the airmass was low for the duration of the observations,
while for the 2010 August 15 and 2010 September 22 transits the airmass increased throughout
the observations and exceeded an airmass of two by the end of the out-of-transit baseline. 
For these latter two observations we noticed that the data quality significantly degraded as the airmass rose above
2.0;
for the analysis that follows we therefore
exclude the data in the out-of-transit baseline with an airmass greater than two for the 2010 August 15 and 2010 September
22 observations.

We observed GJ 1214 in Staring Mode \citep{Devost10} where we use the full WIRCam array 
with its 21\arcmin$\times$21\arcmin \ field-of-view and do not dither for the duration of our observations. 
The only exception to our normal staring mode practice was the aforementioned filter change.
The exposure times and defocus amounts for our various observations were:
3.5 $s$ and 2.0 $mm$ for Ks-band, 4 $s$ and 1.2 $mm$ for the CH$_4$On-filter, and 3.5 $s$ and 1.8 $mm$ for J-band, respectively.
The filter change during our observations
induced an additional overhead of 20 seconds to switch the filter wheel, as well as an additional
60 seconds to take two acquisition images to fine-tune the target position and reinitiate guiding.
We observed in data-cubes of 12 exposures to increase the observing efficiency.
We took 12 exposures (one guide-cube) in a single filter, before
performing the filter change and observing an additional 12 exposures in the other filter.
The resulting duty cycles were: 22\% for our J and Ks-band observations,
and 19\% for our CH$_4$On and J-band observations.

 The data were reduced and aperture photometry was performed on our target star and numerous reference stars. 
We used apertures with radii of 17, 18 and 15 pixels for our Ks, J and CH$_4$On photometry, respectively;
the associated inner
and outer radii for the sky annuli were 21 and 29 pixels for our Ks-band and CH$_4$On photometry,
and 22 and 30 pixels for our J-band photometry.
We preprocess our data using the I'iwi pipeline, designed specifically for WIRCam imaging.
We performed differential photometry on the target star with between 13-25
reference stars for our photometry in our various bands; for further details
on the method we refer the reader to \citet{CrollTrESTwo,CrollTrESThree}.
As can be seen in Figure \ref{FigPoisson},
the out-of-transit photometry after subtraction of the background trend ($\S$\ref{SecResults})
bins down near the Gaussian noise expectation with increasing bin size in all cases.
Our data is therefore not seriously affected by time-correlated red-noise.
We set our errors for our transits equal to the root mean square of the data outside of transit 
after the subtraction of a linear trend; for our J and CH$_4$On photometry on 
2010 August 7 we scale up the errors from the RMS by a factor of $\frac{5}{4}$ as 
this data is of a partial transit only,
and the egress of transit occurs at very high airmass, which we found to be correlated with 
degraded precision with our other data-sets.
The resulting light curves for the various observations are displayed in the top panel of Figure \ref{FigTransits}.

\section{Results}
\label{SecResults}

We fit each of our data-sets with a \citet{MandelAgol02} transit model, with the depth of transit, ($R_p$/$R_{*}$)$^{2}$,
and the best-fit mid-transit time as free
parameters\footnote{We quote the barycentric Julian date in the terrestrial time standard using the routines of \citep{Eastman10}.}.
For several of our GJ 1214 datasets we noticed obvious trends with time after our
differential photometry was performed,
similar to the trends 
noticed in several of our existing WIRCam datasets (e.g. \citealt{CrollTrESTwo,CrollTrESThree,CrollWASPTwelve}).
We cannot rule out that these trends are intrinsic to the target star and could be 
due to, for instance,
long-term stellar variability as a result of rotational modulation.
However,
the frequency with which we observe such trends suggests that most of these trends
are likely systematic in nature. We therefore refer to these trends as background trends,
and we fit our datasets with linear or quadratic backgrounds of the form:
\begin{equation}
B_f = 1 + b_1 + b_2 dt + b_3 dt^2
\end{equation}
where $dt$ is the time interval from the beginning of the observations
and $b_1$, $b_2$ and $b_3$ are fit parameters.
To determine whether a quadratic ($b_3$) term is justified to account
for the background trend,
we calculate the Bayesian information criterion (BIC; \citealt{Liddle07}),
and ensure that the BIC is lower with the inclusion of the quadratic term. That is, the reduction in $\chi^2$ must be
sufficient
to justify the extra degree of freedom. Only the August 15 Ks-band
data warranted a quadratic term ($b_3$).

We employ Markov Chain Monte Carlo (MCMC) fitting as described in \citet{CrollMCMC} and \citet{CrollTrESTwo} using chains 
with 5$\times$10$^6$ steps.
There are four free parameters for each data-set: ($R_p$/$R_{*}$)$^{2}$,
the best-fit mid-transit time, $b_1$, and $b_2$. 
We also fit our data
with the ``residual-permutation'' method \citep{Winn09} where the residuals to the best-fit data are shifted 
and refit, thus preserving the time correlation of 
any red-noise in the data, so as to investigate the impact of time correlated systematics.
We find similar results to our MCMC analysis, although in general the errors are slightly 
larger. The increased size of the errors otherwise is likely
due to low number statistics as in this method for $N$ data-points one can only generate
2$N$-1 light curve permutations ($\sim$200-350 iterations
for each of our data-sets).
For these reasons we quote our MCMC results henceforth.



We employ a quadratic limb-darkening law and obtain our limb-darkening parameters from
\citet{Claret98}
for the J, Ks and CH$_4$On filters -- we adopt their H-band limb-darkening parameters for our 
CH$_4$On filter observations. The \citet{Claret98} limb-darkening parameters are calculated
through fits to the PHOENIX stellar atmosphere models \citep{Hauschildt97a,Hauschildt97b}.
We employ the parameters $c_2$ and $c_4$ from non-linear limb darkening laws, as quoted in Table \ref{TableResults}.
For the input values that we use to generate the limb-darkening parameters
we use approximations to the measured stellar effective temperature ($T_{eff}$ = 3026 $K$), and
log of the stellar surface gravity (log g = 4.991 [CGS units], \citealt{Charbonneau09}), 
of $T_{eff}$ = 3000 $K$ and log g = 5.0.
We adopt the period and ephemeris of GJ 1214b
given in \citet{Bean10}. We use the inclination and semi-major axis to stellar radius 
ratio\footnote{An inclination of i=88.94$^{o}$ and a semi-major axis to stellar radius ratio of a/$R_s$=14.97.}
determined from an analysis from
the \citet{Bean10} white-light curves (J. Bean electronic communication).
All other parameters were fixed at the \citet{Charbonneau09} values.
Our best-fit MCMC transit and background fits for each of our individual transit datasets are listed in the ``Ind.'' rows of 
Table \ref{TableResults}
and displayed in the appropriately coloured lines of Figure \ref{FigTransits}.


 As discussed below ($\S$\ref{SecDiscuss})
we note a small, but significant difference in the transit depths we measure in the J and Ks-bands.
To confirm that this difference is significant, and not due to any uncertainty in the limb-darkening
coefficients\footnote{We also used quadratic limb darkening coefficients
from \citet{LesterNeilson08} and a model without limb-darkening. In both cases our Ks-band
transit depths were significantly deeper than the J-band depths.}
we also fit our three J \& Ks-band transits simultaneously while fitting the two quadratic limb-darkening
coefficients for each band ($c_{2J}$, $c_{4J}$, $c_{2Ks}$, and $c_{4Ks}$).
We place a priori constraints on the limb-darkening coefficients; these priors on the limb-darkening coefficients
are Gaussian with a standard deviation
of 0.05 from the values derived
from \citet{Claret98} (and listed in the ``Ind.'' rows of Table \ref{TableResults} for the J and Ks-bands).
For each of our three transits we fit the Ks and J-band data-sets
with their own background terms (e.g. $b_{1J}$, $b_{2J}$, $b_{1Ks}$, and $b_{2Ks}$ for the 2010 June 27 transit).
We fit for the three mid-transit times of our three transits (to allow for possible transiting timing variations),
but ensure that this value is held in common
between the J and Ks-band transits observed simultaneously.
We fit for the J-band transit depths, but fit the Ks-band transits with a depth that is a consistent 
factor [$(R_{PKs}/R_{PJ})^2$] greater, or less than, the J-band transit observed nearly simultaneously for all three transits.
By utilizing this fitting method, in addition to the methods discussed below in $\S$\ref{SecDiff},
we can directly ascertain how much deeper the Ks-band transits are than the J-band transits.
The advantage of fitting for the depth of each of the J-band transits individually,
rather than fitting them with a consistent depth,
is that this allows for small variations in the J-band depth
that could arise from rotational modulation (as discussed in $\S$\ref{SecSpots}), while still 
directly fitting for the ratio of the Ks to J-band transit depths.
We fit for 24 parameters overall, and the best-fit results
are listed in the rows marked ``Joint'' analysis in Table \ref{TableResults}.

\subsection{WIRCam non-linearity correction}

We also ensured that any difference in the transit depth from the J to Ks-bands was not due to an ineffective 
non-linearity correction.
During the I'iwi preprocessing step,
a non-linearity correction is applied to correct the count levels
for pixels that approach saturation. Near saturation this 
non-linearity correction can be as large as 10\%.
At the maximum count levels recorded in a pixel of the aperture of our target star, GJ 1214, 
during our observations, the detector is well below its saturation level, and the WIRCam
detector is approximately 3-5\% non-linear at these count levels. The vast majority of the pixels in our target
star and reference stars apertures are illuminated to much lower levels and are expected to be non-linear at the 1-3\% level. 
If this non-linearity correction was applied ineffectively then this could cause a systematic offset in our measured
transit depths; although this discrepancy was
expected to be much smaller than the difference in the transit depth from J to Ks
that we measure, we nonetheless demonstrated this was the case by reprocessing and reanalyzing our 
2010 September 22 transit data
in the J and Ks-bands without applying the non-linearity correction.
Any deviations in the pixel count values from the current non-linearity correction, will be more than an order of
magnitude smaller than the effect induced by not applying the non-linearity correction whatsoever.
Not employing the non-linearity corrections, as expected, leads to shallower
transit depths than when the non-linearity
correction is applied. However, the ratio of the transits depths from Ks to J are 
near identical whether the non-linearity correction is,
or is not, applied.
Overall, as this test should create a variation much larger than one due to an ineffective non-linearity
correction, it is safe to conclude that the greater Ks-band than J-band transit depth does not 
arise from the non-linearity correction.

\begin{deluxetable*}{ccccccccccc}
\tablecaption{CFHT/WIRCam near-infrared transit depths of GJ 1214b}
\tabletypesize{\scriptsize}
\tablehead{
\colhead{Date}		& \colhead{Filter} 	& \colhead{Fit}				& \colhead{Mid-Transit Time} 												& \colhead{($R_p$/$R_{*}$)$^{2}$}																	& \colhead{b1}												& \colhead{b2}												& \colhead{b3}								& \colhead{c2}															& \colhead{c4}										 	\\
\colhead{in 2010}	& \colhead{} 		& \colhead{Type\tablenotemark{a}}	& \colhead{(BJD-2450000)} 												& \colhead{(\%)}																			& 													& 													&									& 																& 												\\
}
\startdata
June 27 		& J 			& Ind.					& \JDOffsetZeroGJJ$^{+\JDOffsetPlusZeroGJJ}_{-\JDOffsetMinusZeroGJJ}$							& \TransitDepthPercentAbstractGJJ$^{+\TransitDepthPercentAbstractPlusGJJ}_{-\TransitDepthPercentAbstractMinusGJJ}$							& \cOneGJJ$^{+\cOnePlusGJJ}_{-\cOneMinusGJJ}$								& \cTwoGJJ$^{+\cTwoPlusGJJ}_{-\cTwoMinusGJJ}$ 								& n/a														& \LimbJCTwo 										& \LimbJCFour 											\\
			&			& Joint					& \JDOffsetZeroGJJointAll$^{+\JDOffsetPlusZeroGJJointAll}_{-\JDOffsetMinusZeroGJJointAll}$				& \TransitDepthPercentAbstractGJJointAll$^{+\TransitDepthPercentAbstractPlusGJJointAll}_{-\TransitDepthPercentAbstractMinusGJJointAll}$					& \cOneGJJointAll$^{+\cOnePlusGJJointAll}_{-\cOneMinusGJJointAll}$					& \cTwoGJJointAll$^{+\cTwoPlusGJJointAll}_{-\cTwoMinusGJJointAll}$ 					& n/a														& \ParamEightGJJointAll$^{+\ParamEightPlusGJJointAll}_{-\ParamEightMinusGJJointAll}$ 	& \ParamTenGJJointAll$^{+\ParamTenPlusGJJointAll}_{-\ParamTenMinusGJJointAll}$			\\
June 27 		& Ks 			& Ind.					& \JDOffsetZeroGJKs$^{+\JDOffsetPlusZeroGJKs}_{-\JDOffsetMinusZeroGJKs}$						& \TransitDepthPercentAbstractGJKs$^{+\TransitDepthPercentAbstractPlusGJKs}_{-\TransitDepthPercentAbstractMinusGJKs}$							& \cOneGJKs$^{+\cOnePlusGJKs}_{-\cOneMinusGJKs}$							& \cTwoGJKs$^{+\cTwoPlusGJKs}_{-\cTwoMinusGJKs}$ 							& n/a														& \LimbKsCTwo 										& \LimbKsCFour 											\\
			&			& Joint					& \JDOffsetZeroGJJointAll$^{+\JDOffsetPlusZeroGJJointAll}_{-\JDOffsetMinusZeroGJJointAll}$				& \TransitDepthPercentAbstractKsSixGJJointAll$^{+\TransitDepthPercentAbstractKsSixPlusGJJointAll}_{-\TransitDepthPercentAbstractKsSixMinusGJJointAll}$			& \ParamSixteenGJJointAll$^{+\ParamSixteenPlusGJJointAll}_{-\ParamSixteenMinusGJJointAll}$		& \ParamSeventeenGJJointAll$^{+\ParamSeventeenPlusGJJointAll}_{-\ParamSeventeenMinusGJJointAll}$ 	& n/a														& \ParamTwelveGJJointAll$^{+\ParamTwelvePlusGJJointAll}_{-\ParamTwelveMinusGJJointAll}$ & \ParamFourteenGJJointAll$^{+\ParamFourteenPlusGJJointAll}_{-\ParamFourteenMinusGJJointAll}$	\\
August 7		& J 			& Ind.					& \JDOffsetZeroGJJII$^{+\JDOffsetPlusZeroGJJII}_{-\JDOffsetMinusZeroGJJII}$						& \TransitDepthPercentAbstractGJJII$^{+\TransitDepthPercentAbstractPlusGJJII}_{-\TransitDepthPercentAbstractMinusGJJII}$						& \cOneGJJII$^{+\cOnePlusGJJII}_{-\cOneMinusGJJII}$							& \cTwoGJJII$^{+\cTwoPlusGJJII}_{-\cTwoMinusGJJII}$ 							& n/a														& \LimbJCTwo 										& \LimbJCFour 											\\
			&			& Joint					& n/a															& n/a																					& n/a													& n/a													& n/a														& n/a 											& n/a 												\\
August 7		& CH$_4$On 		& Ind.					& \JDOffsetZeroGJCHFourOn$^{+\JDOffsetPlusZeroGJCHFourOn}_{-\JDOffsetMinusZeroGJCHFourOn}$				& \TransitDepthPercentAbstractGJCHFourOn$^{+\TransitDepthPercentAbstractPlusGJCHFourOn}_{-\TransitDepthPercentAbstractMinusGJCHFourOn}$					& \cOneGJCHFourOn$^{+\cOnePlusGJCHFourOn}_{-\cOneMinusGJCHFourOn}$					& \cTwoGJCHFourOn$^{+\cTwoPlusGJCHFourOn}_{-\cTwoMinusGJCHFourOn}$					& n/a														& \LimbCHFourOnCTwo 									& \LimbCHFourOnCFour 										\\
			&			& Joint					& n/a															& n/a																					& n/a													& n/a													& n/a														& n/a 											& n/a 												\\
August 15		& J 			& Ind.					& \JDOffsetZeroGJJIIIOLAM$^{+\JDOffsetPlusZeroGJJIIIOLAM}_{-\JDOffsetMinusZeroGJJIIIOLAM}$				& \TransitDepthPercentAbstractGJJIIIOLAM$^{+\TransitDepthPercentAbstractPlusGJJIIIOLAM}_{-\TransitDepthPercentAbstractMinusGJJIIIOLAM}$					& \cOneGJJIIIOLAM$^{+\cOnePlusGJJIIIOLAM}_{-\cOneMinusGJJIIIOLAM}$					& \cTwoGJJIIIOLAM$^{+\cTwoPlusGJJIIIOLAM}_{-\cTwoMinusGJJIIIOLAM}$					& n/a														& \LimbJCTwo 										& \LimbJCFour 											\\
			&			& Joint					& \JDOffsetEighteenGJJointAll$^{+\JDOffsetPlusEighteenGJJointAll}_{-\JDOffsetMinusEighteenGJJointAll}$			& \ParamTwentyEightGJJointAll$^{+\ParamTwentyEightPlusGJJointAll}_{-\ParamTwentyEightMinusGJJointAll}$									& \ParamTwentyOneGJJointAll$^{+\ParamTwentyOnePlusGJJointAll}_{-\ParamTwentyOneMinusGJJointAll}$	& \ParamTwentyTwoGJJointAll$^{+\ParamTwentyTwoPlusGJJointAll}_{-\ParamTwentyTwoMinusGJJointAll}$	& n/a														& \ParamEightGJJointAll$^{+\ParamEightPlusGJJointAll}_{-\ParamEightMinusGJJointAll}$ 	& \ParamTenGJJointAll$^{+\ParamTenPlusGJJointAll}_{-\ParamTenMinusGJJointAll}$			\\
August 15		& Ks 			& Ind.					& \JDOffsetZeroGJKsIIQuadraticOLAM$^{+\JDOffsetPlusZeroGJKsIIQuadraticOLAM}_{-\JDOffsetMinusZeroGJKsIIQuadraticOLAM}$	& \TransitDepthPercentAbstractGJKsIIQuadraticOLAM$^{+\TransitDepthPercentAbstractPlusGJKsIIQuadraticOLAM}_{-\TransitDepthPercentAbstractMinusGJKsIIQuadraticOLAM}$	& \cOneGJKsIIQuadraticOLAM$^{+\cOnePlusGJKsIIQuadraticOLAM}_{-\cOneMinusGJKsIIQuadraticOLAM}$		& \cTwoGJKsIIQuadraticOLAM$^{+\cTwoPlusGJKsIIQuadraticOLAM}_{-\cTwoMinusGJKsIIQuadraticOLAM}$		& \cThreeGJKsIIQuadraticOLAM$^{+\cThreePlusGJKsIIQuadraticOLAM}_{-\cThreeMinusGJKsIIQuadraticOLAM}$		& \LimbKsCTwo 										& \LimbKsCFour 										\\
			&			& Joint					& \JDOffsetEighteenGJJointAll$^{+\JDOffsetPlusEighteenGJJointAll}_{-\JDOffsetMinusEighteenGJJointAll}$			& \TransitDepthPercentAbstractKsTwoEightGJJointAll$^{+\TransitDepthPercentAbstractKsTwoEightPlusGJJointAll}_{-\TransitDepthPercentAbstractKsTwoEightMinusGJJointAll}$	& \ParamNineteenGJJointAll$^{+\ParamNineteenPlusGJJointAll}_{-\ParamNineteenMinusGJJointAll}$		& \ParamTwentyGJJointAll$^{+\ParamTwentyPlusGJJointAll}_{-\ParamTwentyMinusGJJointAll}$			& \ParamThirtyGJJointAll$^{+\ParamThirtyPlusGJJointAll}_{-\ParamThirtyMinusGJJointAll}$				& \ParamTwelveGJJointAll$^{+\ParamTwelvePlusGJJointAll}_{-\ParamTwelveMinusGJJointAll}$ & \ParamFourteenGJJointAll$^{+\ParamFourteenPlusGJJointAll}_{-\ParamFourteenMinusGJJointAll}$	\\
September 22		& J 			& Ind.					& \JDOffsetZeroGJJLastOLAM$^{+\JDOffsetPlusZeroGJJLastOLAM}_{-\JDOffsetMinusZeroGJJLastOLAM}$				& \TransitDepthPercentAbstractGJJLastOLAM$^{+\TransitDepthPercentAbstractPlusGJJLastOLAM}_{-\TransitDepthPercentAbstractMinusGJJLastOLAM}$				& \cOneGJJLastOLAM$^{+\cOnePlusGJJLastOLAM}_{-\cOneMinusGJJLastOLAM}$					& \cTwoGJJLastOLAM$^{+\cTwoPlusGJJLastOLAM}_{-\cTwoMinusGJJLastOLAM}$					& n/a														& \LimbJCTwo 										& \LimbJCFour 											\\
			&			& Joint					& \JDOffsetTwentyThreeGJJointAll$^{+\JDOffsetPlusTwentyThreeGJJointAll}_{-\JDOffsetMinusTwentyThreeGJJointAll}$		& \ParamTwentyNineGJJointAll$^{+\ParamTwentyNinePlusGJJointAll}_{-\ParamTwentyNineMinusGJJointAll}$									& \ParamTwentySixGJJointAll$^{+\ParamTwentySixPlusGJJointAll}_{-\ParamTwentySixMinusGJJointAll}$	& \ParamTwentySevenGJJointAll$^{+\ParamTwentySevenPlusGJJointAll}_{-\ParamTwentySevenMinusGJJointAll}$	& n/a														& \ParamEightGJJointAll$^{+\ParamEightPlusGJJointAll}_{-\ParamEightMinusGJJointAll}$ 	& \ParamTenGJJointAll$^{+\ParamTenPlusGJJointAll}_{-\ParamTenMinusGJJointAll}$			\\
September 22		& Ks 			& Ind.					& \JDOffsetZeroGJKsLastOLAM$^{+\JDOffsetPlusZeroGJKsLastOLAM}_{-\JDOffsetMinusZeroGJKsLastOLAM}$			& \TransitDepthPercentAbstractGJKsLastOLAM$^{+\TransitDepthPercentAbstractPlusGJKsLastOLAM}_{-\TransitDepthPercentAbstractMinusGJKsLastOLAM}$				& \cOneGJKsLastOLAM$^{+\cOnePlusGJKsLastOLAM}_{-\cOneMinusGJKsLastOLAM}$				& \cTwoGJKsLastOLAM$^{+\cTwoPlusGJKsLastOLAM}_{-\cTwoMinusGJKsLastOLAM}$				& n/a														& \LimbKsCTwo 										& \LimbKsCFour 											\\
			&			& Joint					& \JDOffsetTwentyThreeGJJointAll$^{+\JDOffsetPlusTwentyThreeGJJointAll}_{-\JDOffsetMinusTwentyThreeGJJointAll}$		& \TransitDepthPercentAbstractKsTwoNineGJJointAll$^{+\TransitDepthPercentAbstractKsTwoNinePlusGJJointAll}_{-\TransitDepthPercentAbstractKsTwoNineMinusGJJointAll}$	& \ParamTwentyFourGJJointAll$^{+\ParamTwentyFourPlusGJJointAll}_{-\ParamTwentyFourMinusGJJointAll}$	& \ParamTwentyFiveGJJointAll$^{+\ParamTwentyFivePlusGJJointAll}_{-\ParamTwentyFiveMinusGJJointAll}$	& n/a														& \ParamTwelveGJJointAll$^{+\ParamTwelvePlusGJJointAll}_{-\ParamTwelveMinusGJJointAll}$ & \ParamFourteenGJJointAll$^{+\ParamFourteenPlusGJJointAll}_{-\ParamFourteenMinusGJJointAll}$	\\
\enddata
\tablenotetext{a}{Fit Type stands for the joint or individual (Ind.) analyses.}
\label{TableResults}
\end{deluxetable*}


\section{Discussion}
\label{SecDiscuss}


\subsection{GJ 1214b's transit depth in the near-infrared}
\label{SecTransits}

\begin{figure*}
\centering
\includegraphics[scale=0.75,angle=270]{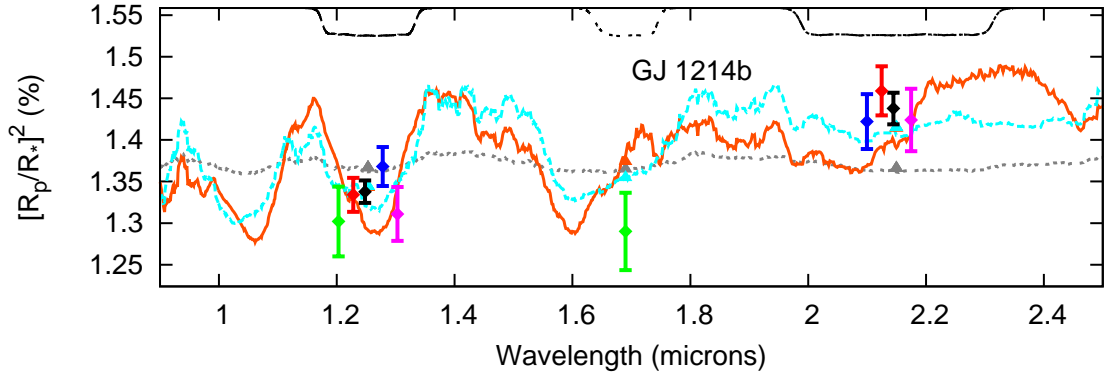}
\caption{	
	WIRCam transit observations of the super-Earth GJ 1214b.
	The WIRCam response functions are displayed inverted at the top of the plot using the black dotted lines at an arbitrary scale; these are from left to right: 
	J-band ($\sim$1.25 $\mu m$), the CH$_4$On filter ($\sim$1.69 $\mu m$), and Ks-band ($\sim$2.15 $\mu m$).
	Our J and Ks-band observations are indicated by the
	red diamonds for our 2010 June 27 observations, the blue diamonds for our
	2010 August 15 observations, and the magenta diamonds for our 2010 September 22 observations.
	Our 2010 August 7 J-band and CH$_4$On filter observations are indicated by the green diamonds.
	We offset our measured J and Ks observations slightly in wavelength for clarity.
	For the Ks and J-band we display the weighted mean and error of the observations in these bands with the black
	diamonds. The orange solid and cyan dashed curves are the \citet{MillerRicciFortney10} GJ 1214b atmospheric models
	for solar metallicity hydrogen/helium dominated atmospheres (with and without methane, respectively). 
	The grey dotted curve
	is the \citet{MillerRicciFortney10} models for an H$_{2}$O/steam atmosphere. 
	We integrate the \citet{MillerRicciFortney10} atmospheric models over the WIRCam response functions
	and display these values in the appropriately coloured solid triangles.
	}
\label{FigGJ1214}
\end{figure*}

We display our best-fit transit depths in Figure \ref{FigGJ1214} and Table \ref{TableResults}.
The J-band transit depths are largely consistent with one another and are also consistent, or at most insignificantly shallower,
than the depths reported by \citep{Charbonneau09} and \citet{Bean10} in the optical and very near-infrared.
The Ks-band transits also display similar depths to one another.
However, the Ks-band transits appear to be deeper than the J-band transits; 
this is a small effect, but is clearly visible in the bottom panel of Figure \ref{FigTransits} where
we present the residuals of our observations from the best-fit J-band transit depths observed nearly simultaneously.

	The CH$_4$On transit depth, on the other hand, appears to have a similar transit depth to the 
the J-band transit observed simultaneously on 2011 August 7.
As this is a partial transit only, and as much of the transit and the egress of transit occurs at very high airmass, caution
is warranted in any robust comparison of the J to CH$_4$On transit depth and to other wavelengths.

By combining all the Ks-band and J-band transits, we find the
weighted means and the associated errors on the transit depths are 
($R_{pJ}$/$R_{*}$)$^{2}$=\GJJBandCombinedTransitDepthPercent $\pm$\GJJBandCombinedTransitDepthPercentMinusPlus\% for J-band,
and ($R_{pKs}$/$R_{*}$)$^{2}$=\GJKsBandCombinedTransitDepthPercent $\pm$\GJKsBandCombinedTransitDepthPercentMinusPlus\% for Ks-band.
As we only have one partial transit of GJ 1214 in the CH$_4$On filter,
the depth in that band is simply the value from the 2010 Aug. 7 transit:
($R_{pCH_4On}$/$R_{*}$)$^{2}$=\TransitDepthPercentAbstractGJCHFourOn$^{+\TransitDepthPercentAbstractPlusGJCHFourOn}_{-\TransitDepthPercentAbstractMinusGJCHFourOn}$\%.
We determine the error on the weighted mean of our J-band and Ks-band points
by determining the weighted error of all our observations
in that particular 
band and then scaling that error upwards by a factor of $\zeta$. 
To determine $\zeta$ we calculate the $\chi^{2}$ of all our data
in a single band compared to a model with a consistent transit depth
equal to the 
weighted mean of the transit depths
in that band; we then scale up the errors to ensure the reduced $\chi^{2}$ is equal to 
one\footnote{\citet{Andrae10} notes that there are several hidden assumptions one should be careful
to address when applying reduced $\chi^2$ to one's data;
we feel the method we apply here should be useful nonetheless
as a first-order approximation to indicate the appropriate size of the weighted error bars}.
The Ks-band data-points are consistent with one another, so only the J-band errors are scaled upwards.
The resulting value is $\zeta_J$=\ZetaJ \ for the J-band photometry, so this suggests that both the Ks and J-band
weighted errors are already appropriately sized, or close to it.
Overall, this analysis suggests that our Ks-band and J-band transit depths are inconsistent with
one another; the Ks-band transit depth is deeper than the J-band 
depth with \GJKsJSigmaDiff$\sigma$ confidence.

\subsection{The effect of stellar spots on transit observations of GJ 1214b}
\label{SecSpots}

\citet{Charbonneau09} reported that GJ 1214 is an active star and displays longer term variability with a period of several weeks
at the 2\% level in the MEarth
bandpass. More recently, \citet{Berta10} presented and analyzed MEarth photometry of GJ 1214 from 2008 to 2010
and observed long-term photometric variability at the 1\% level with a period of approximately 50 days.
This variability is presumably due to rotational modulation from spots rotating in and out of view.
As the long-term photometric monitoring presented in \citet{Berta10} ends in 2010 July (in the midst of the observations we present here)
we assume the more conservative
limit of 2\% variability for our calculations henceforth.

Transit observations obtained at different
epochs may show small differences in the transit depth due to rotational modulation arising from both occulted
and unocculted spots \citep{Czesla09,Berta10,Carter11}. 
In the case of unocculted spots, if the 2\% observed
rotational modulation represents the full range from a spotted to unspotted 
photosphere\footnote{There is no reason to expect we ever observe a hemisphere of the star free of spots altogether, and indeed
the analysis of \citet{Carter11} and \citet{Berta10} suggest we very well may not, which would lead to larger transit
depth differences from epoch to epoch. Exact analytical expressions are available in \citet{Carter11}.}
then we may expect 
measurements of the transit depth of GJ 1214b will vary by as much as 
0.03\% of the stellar signal in the MEarth 
bandpass (assuming an unspotted transit depth of 1.35\%)
from observations taken at epochs spanning
the maximum and the minimum of the observed rotational modulation.
On the other hand, occulted spots will cause small brightening events during the transit that may lead one to underestimate the true transit depth.
Thus, for transit depth measurements obtained at different epochs, such as our own, it is possible that the variability
caused by rotational modulation
creates
small differences in the measured depths for observations taken at different epochs.

	However, for our data obtained nearly simultaneously in two 
different bands, the effect of spots should be reduced for the following reasons. First of all,
as the stellar rotation period of GJ 1214 appears to be much longer \citep{Charbonneau09,Berta10} than the 
$\sim$1 hour duration of the transit, 
the spot pattern should be essentially static during a single transit.
Secondly, even if the star is very spotted during our own observations, the difference between the transit depths measured 
nearly simultaneously in
our two bands will be minute.
This small difference will arise due to the differing ratio of the Planck function of the spot and the star
due to their different temperatures; however, this
difference will be muted as we move into the near infrared. 
For instance, assuming that GJ 1214 has spots 500 K cooler -- a value supposedly
consistent with another M4.5 dwarf \citep{Zboril03} --
than GJ 1214's $\sim$3000 K effective temperature \citep{Charbonneau09},
the 2.0\% variability due to rotational modulation in the MEarth bandpass ($\sim$780 $nm$), will translate 
into 1.5\%, and 1.0\% variability in J-band and 
Ks-band, respectively. 
Assuming the unspotted transit depth
is 1.350\% in these bands then the maximum transit depths from unocculted spots,
which would results from measurements
at the minimum flux of the observed rotational
modulation, would be 1.369\% in J and 1.364\% in Ks-band.
The variation in the transit depth between the J and Ks-bands that we observe is both much larger
than this predicted effect due to starspots, and also would serve
to create a deeper transit in J-band, rather than Ks-band; we, of course, observe the opposite phenomenon.

A deeper Ks-band than J-band transit could arise from spots along the transit chord that are occulted during the observations.
Occultation of spots by a planet will create anomalous brightenings
during transit, as has been observed for the transiting planets
HD 189733b \citep{Pont07}, TrES-1b \citep{Rabus09,Dittmann10}, and more recently for GJ 1214b \citep{Bean10,Berta10,Carter11}.
However, due to the near-simultaneous nature of our photometry, it is unlikely that occulting spots could account for the
variation in Ks to J-band transit depth that we observe, again
due to the small difference in the Planck functions of the spot relative to the star between
our two bands. For instance, occulting a spot 500 K cooler than the surrounding photosphere that is 30\% of the planetary size,
will lead to transits that are 0.07\% shallower for the duration of the occultation than the presumed 1.35\% transit
depth in J-band. 
However the Ks-band transit depth will also be 0.05\% shallower; the 0.02\% relative difference between the two bands expected from a spot
occultation
is much less than the observed transit depth difference we observe. 
Lastly, as spot occultations will lead to overall shallower transit depths, one would require the true transit depth of GJ 1214b to be
deeper than that observed in our own J-band observations and in the optical and very near-infrared \citep{Charbonneau09,Bean10}.
For these reasons we find it unlikely that the transit depth difference we observe arises from occulted or unocculted spots.

\subsection{A larger transit depth in Ks-band than J-band}
\label{SecDiff}

\begin{figure}
\centering
\includegraphics[scale=0.60, angle = 270]{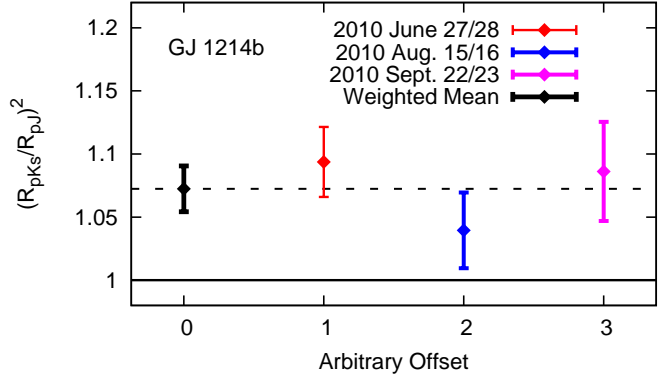}
\caption{	The measured Ks-band transit depth divided by the measured 
		J-band transit depth for the ``Ind.'' analysis of our various observations (see the legend).
		The horizontal dotted line denotes the weighted mean
		of the Ks divided by J-band transit depth, $(R_{PKs}/R_{PJ})^2$.
		The solid horizontal solid line denotes the value if the Ks-band transits were the same depth as the 
		J-band transits. 
		}
\label{FigDivision}
\end{figure}

\begin{figure}
\centering
\includegraphics[scale=0.60, angle = 270]{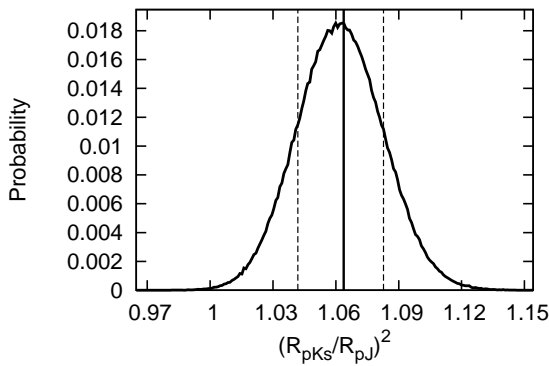}
\caption{	The measured Ks-band transit depth divided by the measured 
		J-band transit depth from our ``Joint'' analysis.
		The vertical solid line denotes the
		best-fit value, while the 
		dotted vertical lines indicate
		the 68\% credible regions.
		}
\label{FigJoint}
\end{figure}

 Due to the aforementioned possible variations in the transit depths from epoch to epoch
induced by spots, a more straightforward
method to compare the depths of transits in our bands is to directly compare 
the depth of the transit in one band to the depth obtained simultaneously in another - that is $(R_{PKs}/R_{PJ})^2$ for
each one of our transits observed simultaneously in the Ks and J-bands.
We ignore our data observed on 2010 August 7 for this analysis, as GJ 1214 was observed
in the J-band and the CH$_4$On filter, rather than in J and Ks.

We attempt to measure the fraction that the Ks-band transits are deeper than the J-band transits, $(R_{PKs}/R_{PJ})^2$, 
by two methods. In the first method, we 
display the best-fit MCMC transit depth of our Ks-band photometry divided by the J-band transit depth from our individual 
analysis (``Ind.'') in Figure \ref{FigDivision}.  
The associated errors displayed in this figure are
generated by propagating
through the associated errors on the individual best-fit MCMC transit depths
as displayed in Table \ref{TableResults}.
The weighted mean and error of these data indicates that the Ks-band transits 
are deeper than the J-band transits
by a factor of: $(R_{PKs}/R_{PJ})^2$=\GJKsOverJCombined$\pm$\GJKsOverJCombinedMinusPlus.
The associated errors do not need to be scaled up, as the reduced $\chi^{2}$ is near one for a comparison
of our three $(R_{PKs}/R_{PJ})^2$ data-points as compared to the weighted mean
of these observations; specifically $\chi^{2}$=\GJKsOverJChiSquared, which is reasonable
given the two degrees of freedom\footnote{We again note there are several hidden
assumptions one should be aware of when applying reduced $\chi^2$ as documented
in $\S$\ref{SecTransits}.}
By analyzing the individual transits, our Ks-band photometry displays a deeper transit depth than our
J-band photometry observed nearly simultaneously with a confidence in excess of \GJKsOverJSigmaDiff$\sigma$.

In the second method we use the ``Joint'' MCMC analysis described above where we simultaneously 
fit the three three data-sets that observe in Ks-band
and J-band simultaneously.
We fit each J-band transit with an independent transit depths, but assume all
the Ks-band transits were a consistent factor deeper
(or shallower) than the J-band transits observed nearly simultaneously, $(R_{PKs}/R_{PJ})^2$.
We also fit the limb-darkening parameters after applying a Gaussian a priori assumption as described in $\S$\ref{SecResults}.
We display this ratio for our ``Joint'' analysis in Figure \ref{FigJoint}; 
we measured this fraction as 
$(R_{PKs}/R_{PJ})^2$= \ParamFifteenGJJointAll$^{+\ParamFifteenPlusGJJointAll}_{-\ParamFifteenMinusGJJointAll}$.
This distribution is not a perfect Gaussian, and therefore, according to this method
our Ks-band transits are deeper than our J-band transits with
greater than \DiffSigmaIntJointKsJGJJointAll$\sigma$ confidence.

Both methods return reasonably similar results, and argue for 
a deeper Ks-band transit depth than J-band transit depth with a confidence in excess
of \DiffSigmaIntJointKsJGJJointAll$\sigma$.
We quote the value of $(R_{PKs}/R_{PJ})^2$
from our ``Ind.'' results henceforth.

\subsection{WIRCam transit depths suggest a low mean molecular weight}
\label{SecModels}

\begin{figure*}
\centering
\includegraphics[scale=0.75,angle=270]{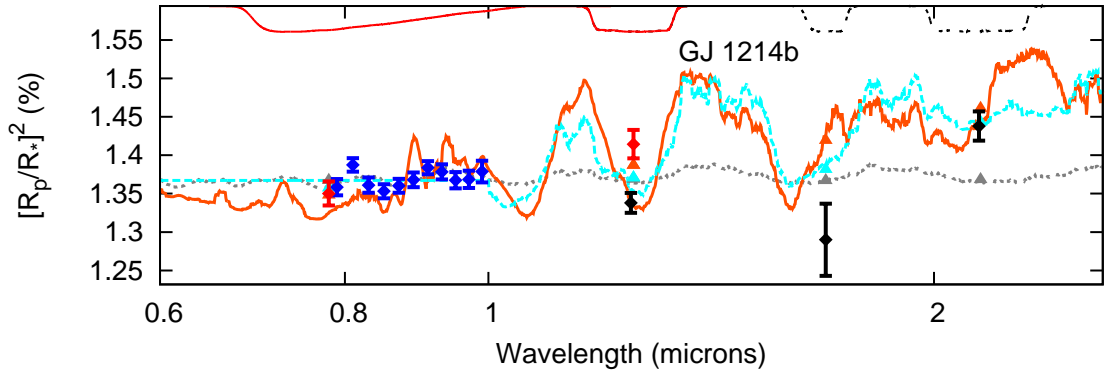}
\caption{	
	Transit observations of the super-Earth GJ 1214b.
	The response functions are 
	displayed inverted at the top of the plot at arbitrary scale.
	These are from left-to-right the MEarth ($\sim$0.78 $\mu m$), and Kitt-peak J-band ($\sim$1.25 $\mu m$) 
	displayed in the red-curves, and the WIRCam 
	J-band ($\sim$1.25 $\mu m$), CH$_4$On filter ($\sim$1.69 $\mu m$), and Ks-band 
	($\sim$2.15 $\mu m$) displayed in the black dotted curves. 
 	The Kitt-peak J-band bandpass overlaps the CFHT J-band bandpass.
	The VLT/FORS2 response functions are not included,
	as they essentially overlap with the points. 
	The weighted mean of our J-band, Ks-band and CH$_4$On filter observations are 
	displayed by the black diamonds. The MEarth \citep{Charbonneau09},
	and the Kitt-peak \citet{Sada10} transit depth
	are displayed with the red diamonds, while the VLT/FORS2 points \citet{Bean10} 
	are displayed with blue diamonds.
	We display two \citet{MillerRicciFortney10} GJ 1214b atmospheric models 
	(a water-world model [grey dotted curve], and a solar-metallicity model [orange solid curve]);
	the other model we present in the cyan dashed curve is
	a \citet{MillerRicciFortney10} solar-metallicity no methane model at wavelengths longer than one micron, 
	while at shorter wavelengths we arbitrarily cut off the predicted absoprtion to simulate the impact of putative hazes.
	We integrate these atmospheric models over the WIRCam response functions
	and display these values in the
	appropriately coloured solid triangles.
	We caution that this comparison does not correct the depths
	for possible variability due to rotational modulation
	and thus any comparison should be considered illustrative, rather than definitive.
	}
\label{FigAll}
\end{figure*}

The transit depths from our CFHT/WIRCam photometry
suggests that GJ 1214b should have a large scale height, low mean molecular weight
and thus a hydrogen/helium dominated atmosphere.
These conclusions arise from the fact that our Ks-band transit depths are deeper than our J-band transit depths by a factor
of 
$(R_{PKs}/R_{PJ})^2$=\GJKsOverJCombined$\pm$\GJKsOverJCombinedMinusPlus.
This corresponds to a relative change in the radius of GJ 1214b from the Ks-band to the J-band of \RadiusKsIncreaseGJKsOverJ\%, or
$\sim$\KilometersRadiusKsIncreaseRoundGJKsOverJ \ $km$ compared to GJ 1214b's radius of $\sim$17070 $km$.
Using its equilibrium temperature
of $T_{eq}$$\sim$560 $K$ (the value obtained assuming zero bond albedo)
this amounts to absorption increasing the radius
of GJ 1214b by $\sim$2 atmospheric scale heights for a hydrogen gas envelope (H$_2$).
As, the molecular weight of $H_2O$ is approximately nine times greater than hydrogen gas,
a spectral feature this prominent assuming a water world composition would require 
an increase in the planetary radius of $\sim$20 atmospheric scale heights.
This suggests that an atmosphere
dominated by light elements is much more probable than one composed of heavier elements.

The change in transit depth due to a change in planetary radius between the line, $R_{PL}$,
and the continuum, $R_{PC}$, 
can be related to the scale height, $H$,
and the opacities in the absorption line, $\kappa_{l}$, and the continuum, $\kappa_{c}$;
this value can be approximated by the ratio of the opacities multiplied by the
area of an annulus one scale height thick relative to that of the stellar disk:
\begin{equation}
(R_{PL}/R_{*})^2-(R_{PC}/R_{*})^2 = \frac{2 \pi R_{P} H}{\pi R_*^2} \ln(\kappa_{l}/\kappa_{c}).
\end{equation}
\citep{Brown01}.
To cause the observed transit depth difference in a hydrogen gas atmosphere would require 
line opacity marginally greater than that of the continuum ($\kappa_{l}/\kappa_{c}$$\sim$8);
the water-world composition would
require an opacity that is unrealistically larger 
than that of the continuum ($\kappa_{l}/\kappa_{c}$$\sim$2$\times$10$^{8}$).
Therefore from our CFHT/WIRCam observations, one would expect that the atmosphere
of GJ 1214b must have a low mean molecular weight, a
large scale height, and thus an atmosphere dominated by hydrogen and/or helium.

We compare our 
data to the \citet{MillerRicciFortney10} atmospheric models of GJ 1214b in 
Figure \ref{FigGJ1214}. We first employ the most conservative
scenario and assume that due to rotational modulation, or other systematic errors,
that we cannot directly compare our measured transit depths to the \citet{Charbonneau09} or \citet{Bean10} depths.
We thus scale-up or down the predicted absorption of the \citet{MillerRicciFortney10} models
by a multiplicative factor, analogous to an increase or decrease of the squared ratio of the planetary to stellar radius,
to produce the best-fit (minimum $\chi^{2}$) compared to our observations.
We display the \citet{MillerRicciFortney10} water-world model 
(an $H_2O$ dominated world) in the dotted
grey line.
We compare to two hydrogen-helium dominated atmospheres; the first has 
solar-metallicity (orange dot-dash line) while the second has solar metallicity
but does not feature methane (cyan dot-dashed line).
We integrate these models over the WIRCam response functions and calculate the associated $\chi^{2}$
of our data compared to the model.
Although, we cannot strongly
differentiate between
hydrogen-helium envelopes with solar metallicity ($\chi^2_{solar}$=\ChiSolar) 
from those without methane ($\chi^2_{no-methane}$=\ChiNoMethane), 
the water-world composition is disfavoured ($\chi^2_{H_2O}$=\ChiHTwoO) by greater than 
\NSigmaModels$\sigma$ for our seven degrees of freedom. Other high mean molecular weight models (e.g. the \citet{MillerRicciFortney10}
CO$_2$-dominated or H$_2$O/CO$_2$-mixture atmospheres) are disfavoured with similar confidence.
The support for the hydrogen/helium composition arises from the observed increased absorption in Ks-band as opposed to our J-band observations.
The CH$_4$On filter is nominally 1$\sigma$ discrepant from the observed models; it is unclear at this present time whether this discrepancy
is physical or simply due to low signal-to-noise.

\subsection{Comparison to observations at other wavelengths}

	Arguably, completely excluding the constraints imposed 
by transit depth observations at other wavelengths, 
because of the effects of spots or systematic effects,
is unnecessarily conservative.
Therefore, we also compare the weighted means of our transit depth measurements and the \citet{MillerRicciFortney10} models to the 
\citet{Charbonneau09} depth measurement in the MEarth bandpass, the J-band measurement of \citet{Sada10}, and the \citet{Bean10} measurements
from 0.78-1.0 $\mu m$ in Figure \ref{FigAll}. We do not attempt to make a correction for the possibly
variable spot activity between the various epochs at which the data were obtained; as discussed above in $\S$\ref{SecSpots}
rotational modulation can be expected to induce spurious transit depth variations as large as 0.03\%
of the $\sim$1.35\% transit depth of GJ 1214b near 1 $\mu m$, and will cause smaller variations at longer wavelengths. 
Also,
a small discrepancy will be induced by the fact that \citet{Sada10} and \citet{Charbonneau09} transit depth measurements were 
produced with the original \citet{Charbonneau09} estimates of the inclination and other parameters for this system, rather
than the values derived from the \citet{Bean10} white light photometry that we use here and that have been applied
to the \citet{Bean10} spectrophotometry.
For these reasons we caution that this comparison should be
considered illustrative, rather than definitive. 

 Of the original \citet{MillerRicciFortney10} atmospheric models the heavy mean molecular weight models, such as 
the water-world composition model ($\chi^2_{H_2O}$=\ChiHTwoOEVERYTHING), are highly favoured over the 
low mean molecular weight compositions ($\chi^2_{solar}$=\ChiSolarEVERYTHING \ 
for the solar metallicity hydrogen/helium dominated
envelope).\footnote{As explained above in $\S$\ref{SecModels}, we scale the radii of the planet in the models up or down to achieve
the minimum $\chi^2$ compared to the data.} The water-world composition is favoured over the solar metallicity hydrogen/helium
dominated model with more than 
5$\sigma$ confidence. This is unsurprising, as it is largely the conclusion of the \citet{Bean10} paper, and results
from the high precision of 
the VLT/FORS2 spectrophotometry and the lack of observed spectral features in the very near-infrared.
The \citet{Bean10} paper excludes the expected methane and water spectral absorption
features from a hydrogen/helium dominated atmosphere from 0.78 - 1.0 $\mu m$ with high confidence. 

	Another possibility to explain the lack of observed spectral features are high altitude hazes \citep{Fortney05}
in the atmosphere of GJ 1214b that could mute the spectral 
features at shorter wavelengths. We discuss this possibility further below ($\S$\ref{SecHazes}). We thus also compare
our observations to a no methane model where we cut off the absorption below 1 $\mu m$ and set it equal to a nominal value of 1.35\%. 
At wavelengths greater than 1 $\mu m$, the values are identical to the no methane model.
This abrupt transition is not intended to be physical,
but simply illustrative of the impact hazes could have on the expected transmission spectrum at shorter wavelengths\footnote{Depending
on the size of the particles one would expect the observed transit radius to increase at
very short wavelengths due to Rayleigh scattering or have a more complicated behaviour due to Mie scattering.}.
This model has the lowest
$\chi^{2}$ ($\chi^2_{hazy-hydrogen}$=\ChiHazeEVERYTHING) of any of the models as
it is able to address the lack of observed
spectral features in the \cite{Bean10} spectrophotometry, and our deeper Ks-band depth compared to our J-band
depth\footnote{While this paper
was in the late stages of revision, \citet{Desert11}
presented Spitzer 3.6 and 4.5 $\mu m$ channel observations of the transit of GJ 1214b.
They find similar transit depths in these bands to those found by \citet{Bean10} and \citet{Charbonneau09}
in the very near-infrared. If methane is present in the atmosphere of GJ 1214b
it should cause increased absorption in the 3.6 $\mu m$ channel; as this is not observed,
if the atmosphere of GJ 1214b is hydrogen/helium dominated,
the combination of all the observations to date argues in favour of the hazy hydrogen/helium dominated model without methane.}.
We note that the improvement in the $\chi^{2}$ of this haze model to the water-world model is not significant, and both remain
leading candidates to explain all the observations of GJ 1214 to date, as discussed below.


\subsection{Possible atmospheric compositions of GJ 1214b}
\label{SecHazes}

	All the transmission spectroscopy observations of GJ 1214b to date
could be explained by a high mean molecular weight atmosphere if our deeper Ks-band transit is simply an outlier.
If GJ 1214b's atmosphere does have a high mean molecular weight, a
water vapour atmosphere is a leading possibility. This water-world scenario will remain a viable candidate until 
observations are performed to either
confirm that our Ks-band transit depths are indeed deeper, or spectral features are detected at other
wavelengths.

	A scenario that would qualitatively explain all the observations to date,
is a hydrogen/helium dominated atmosphere with thick hazes that would mute
the presence of spectral features arising from shorter wavelengths due to scattering. This haze layer would have
to be at high altitudes,
and low pressures ($<$200 mbar),
to effectively mute the expected spectral features that would arise from being able to stare deep into the atmosphere
of the planet in opacity windows in the very near-infrared.
The efficiency of scattering diminishes for wavelengths longer than the approximate particle
size \citep{HansenTravis74}.
The haze particles could not be much smaller than sub-micron size to account for the 
the lack of observed
spectral features in the \citet{Bean10} spectrophotometry in the very near-infrared. 
Due to the expected size of these putative haze particles, 
shorter wavelength optical observations 
would not be expected to show simply the monotonic increase in planetary radius expected from a Rayleigh scattering
signal, but instead the more complicated transmission spectrum signal of Mie scattering (see for example \citealt{Lecavelier08}). 


Such a haze or cloud layer is certainly not inconceivable. A cloud deck or haze has been reported
to mute the optical transmission spectrum from 0.29 - 1.05 $\mu m$ of the hot Jupiter 
HD 189733b \citep{Pont08,Sing11}; in 
the infrared HD 189733b may have absorption features with an opacity even greater than those that result from the 
haze at those wavelengths \citep{Desert10}.
The hazes of Jupiter and Titan may be
other suitable analogies. Titan has a haze layer that is optically thick 
in the optical, but has transparent windows as one moves into the 
near-infrared \citep{Tomasko08,Griffith93}. The opacity of hazes on Jupiter are high 
at short optical wavelengths, but are much smaller as 
one moves into longer optical wavelengths
and into the near-infrared \citep{Rages99}.

A potential culprit for the particle causing this haze is a hydrocarbon
derived from the photochemical destruction of methane \citep{Moses05,Zahnle09}. Methane is 
found in the atmospheres of all the solar system's giant planets, as well as Titan.
The end product of the breakdown of methane are higher order hydrocarbons that condense as
solids \citep{Rages99}.  Since GJ1214b should be relatively cool ($T_{eq}$$\sim$500K) its atmospheric carbon
inventory could feature abundant methane, like Jupiter.  Particulates with a relatively small mixing
ratio can have important effects at the slant viewing geometry appropriate for transits \citep{Fortney05},
and thus hydrocarbons in a high altitude haze
are one possible explanation the lack of observed features in the \citet{Bean10} spectrophotometry.

 We lastly note, that the actual spectral features of GJ 1214b, whether its atmosphere is hydrogen/helium dominated or not, could be
more complicated and thus very different than the \citet{MillerRicciFortney10} models predict. One such reason could be the impact of
non-equilibrium chemistry, which will be explored in a forthcoming publication (Miller-Ricci Kempton et al. in prep.).

	Clearly further observations are required to differentiate between these scenarios and determine the true 
atmospheric makeup of GJ 1214b.

\subsubsection{An opacity source at $\sim$2.15 $\mu m$}

	The increased transit depths we note in our Ks-band observations, argue for an opacity source near
$\sim$2.15 microns that is causing
increased absorption along the limb of the planet. One such, possible opacity source is methane, which is predicted
to cause absorption from $\sim$2.2 to $\sim$2.4 $\mu m$
in the \citet{MillerRicciFortney10} 
hydrogen-helium dominated atmospheric models.
Although our Ks-band transit depth is qualitatively consistent with this methane spectral absorption feature, we
note that the \citet{MillerRicciFortney10} hydrogen/helium
model with solar metallicity but without methane, provides a near-identical goodness-of-fit
as compared to the solar metallicity model with methane.
This is because the methane absorption feature is present at the red-edge of our Ks-band only; on the blue edge of the
Ks-band the hydrogen/helium model with methane actually features less absorption than the model without methane, so overall
the predicted transit depth is similar whether methane is or is not present according to the \citet{MillerRicciFortney10}
prediction.
Also, greater concentrations of methane in the atmosphere are not expected to cause 
increased absorption at these wavelengths. In the no methane model, the increased Ks-band absorption
compared to J-band results from water opacity. Thus both water and methane remain viable candidates
for this increased absorption. Both
these molecules should
also lead to
spectral features from 0.78 - 1.0 $\mu m$ that have been ruled out by the \citet{Bean10}
spectrophotometry at high confidence.
Thus for these chemicals to remain viable opacity sources, one requires the presence of hazes in the atmosphere of
GJ 1214b, or that the spectral features are more
complicated than the \citet{MillerRicciFortney10} models predict. We encourage further modelling to elucidate whether
this $\sim$2.15 $\mu m$ feature is due to methane, water or another opacity source.


\subsection{Consequences of a hydrogen/helium dominated atmosphere}

A hydrogen/helium dominated atmosphere on GJ 1214b would be expected to undergo significant
hydrodynamic escape. Therefore, if GJ 1214b is hydrogen/helium dominated then it 
may have lost or is losing a significant fraction
of its gaseous envelope. \citet{Charbonneau09} and \citet{RogersSeager10}
predicted that if GJ 1214b's atmosphere is dominated by hydrogen gas then it will lose on the order
of $\sim$10$^{9}$ $g$$s^{-1}$, or $\sim$0.02 $M_{\earth}$ on a 4 Gyr
timescale\footnote{\citet{Charbonneau09} report a 3-10 Gyr age for this system}.
As its host star may have been more active earlier in its life, and thus brighter in the ultraviolet, its
cumulative mass loss may be higher.
Thus, either if GJ 1214b's hydrogen/helium envelope is primordial or due to outgassing
we may expect it to have lost and will be losing a non-negligible fraction of its atmosphere.

Also, \citet{Carter11} have pointed out that the radius of the star, GJ 1214,
that one obtains from stellar evolutionary models is very different than the radius one obtains from an analysis
of the light curve. As a result the density of GJ 1214b varies accordingly, from one where a significant gaseous atmosphere
is likely, to a much higher density where one would only expect a thin gaseous atmosphere on top of a solid terrestrial planet.
Our increased Ks-band transit depth argues in favour of the lighter density and the stellar radius suggested by fits to 
light curve parameters.
 
\subsubsection{Constraints on GJ 1214b's bulk composition}
\label{SecCore}

 If the atmosphere of GJ 1214b is hydrogen/helium dominated then this allows us 
to place a constraint on the planet's bulk composition, namely 
its core, mantle and possibly its ice layer \citep{RogersSeager10,Nettelmann10}. 
This is because the lighter atmospheric composition of a hydrogen/helium atmosphere,
compared to  for instance a water-world composition,
requires a heavier interior composition of silicates, ferrous material or ices 
to compensate in order to fit the observed mass and radius -- and thus density --
constraints of \citet{Charbonneau09}. 
We compare our results to two numerical models \citep{RogersSeager10,Nettelmann10}
that attempt to determine the range of realistic
bulk compositions that agree with the observed mass and radius constraints. 
Regardless of whether the planet's hydrogen/helium
envelope is primordial or due to outgassing, it is expected that this atmospheric layer will be a small percentage ($\sim$5\%)
of this planet's total mass \citep{RogersSeager10,Nettelmann10}.
\citet{Nettelmann10} and \citet{RogersSeager10} predict, under their hydrogen/helium atmosphere scenarios,
that a wide range of core/mantle masses is still viable (from a few percent to 
$\sim$99\% of the mass of GJ 1214b). The higher core/mantle
masses result from a planet with very little water, while for the lower masses it 
would entail a massive interior water/ice layer.
\citet{Nettelmann10} suggest that if this planet's atmosphere is dominated by hydrogen and helium then one
can place an upper limit on the water to rock ratio of approximately six-to-one;
the true value of this quantity is expected to be much lower, and thus the core mass is expected to make-up at least 
$\sim$14\% of the planet's mass and likely much more \citep{Nettelmann10}.

%
%




\section{Conclusions}

We report observations of four transits of GJ 1214b using WIRCam on CFHT. We observed nearly simultaneously in J-band
and in Ks-band for three of the transits, and in J-band and the CH$_4$On filter in another. 
Our best-fit J-band transit depth is consistent with the values obtained in the optical and very near-infrared: 
$(R_{PJ}/R_{*})^2$=\GJJBandCombinedTransitDepthPercent $\pm$\GJJBandCombinedTransitDepthPercentMinusPlus\%. Our Ks-band transit is
deeper: $(R_{PKs}/R_{*})^2$=\GJKsBandCombinedTransitDepthPercent $\pm$\GJKsBandCombinedTransitDepthPercentMinusPlus\%.
Our J and Ks-band transit depths are inconsistent at the 
\GJKsJSigmaDiff$\sigma$ level.
The impact of spots and rotational modulation on the transit depths we
observe should be small; nevertheless spots will induce small changes in the transit depths we measure from epoch to epoch, and
as a result a better metric to quantify our observations may be
the factor that the Ks-band transits are deeper than the J-band transits observed simultaneously. 
Our Ks-band transits display a deeper depth than our J-band transits by a factor of
$(R_{PKs}/R_{PJ})^2$=\GJKsOverJCombined $\pm$\GJKsOverJCombinedMinusPlus.
We thus detect increased transit depths in our broadband Ks-band ($\sim$2.15 $\mu$m) as compared to J-band ($\sim$1.25 $\mu$m) with 
\GJKsOverJSigmaDiff$\sigma$ confidence.
The difference in transit depth between the two bands that we measure is best explained due to a spectral absorption feature
from the atmosphere of GJ 1214b; the only way to get a spectral absorption feature this prominent is if the atmosphere of GJ 1214b has a 
large scale height, low mean molecular weight and is thus hydrogen/helium dominated. 
Water or methane are possible opacity sources to explain this absorption. 
If GJ 1214b's atmosphere is hydrogen/helium dominated
a range of core/mantle masses and ice layers is still viable, but
the planet must have a large rocky core/mantle interior to its gaseous envelope. In this case, our
increased Ks-band transit depth would be the first detection of a spectral feature in a super-Earth atmosphere, and 
GJ 1214b would
best be described as a 
mini-Neptune.


	However, when combining our observations with other observations of GJ 1214b, most specifically
the lack of spectral features observed in the \citet{Bean10} VLT/FORS2 spectrophotometry
from 0.78 - 1.0 $\mu m$, the atmospheric composition of GJ 1214b is less clear. There are several leading possibilities.
One possibility remains that the atmosphere of GJ 1214b is 
water-vapour dominated and our increased Ks-band transit depth is simply an outlier; 
our increased Ks-band transit depth will have 
to be reconfirmed 
or spectral features at other wavelengths will have to be detected 
before this scenario can be confidently ruled out.
The possibility that is arguably the most consistent with all the observed data to date,
is that GJ 1214b has a hydrogen/helium dominated
atmosphere with a haze layer at high altitude
consisting of particles that can be no smaller than approximately sub-micron in size;
such a scenario would explain the lack of observed spectral
absorption features in the very near-infrared in the \citet{Bean10} spectrophotometry
and our own increased Ks-band transit depth if there is an opacity source at $\sim$2.15 $\mu m$. 
Lastly, we note that the true spectrum of GJ 1214b could be more complicated than our models predict for a variety of reasons - 
one such possibility is the importance of non-equilibrium chemistry in GJ 1214b's atmosphere which would alter
GJ 1214b's predicted transmission spectrum (Miller-Ricci Kempton et al. in prep.). 

	Clearly, further observations are required to precisely determine the nature of GJ 1214b's atmosphere.
We encourage further observations
to confirm our increased Ks-band transit depth. We plan to use CFHT/WIRCam to reobserve the transit of GJ 1214b
in Ks-band on several occasions
in the Spring/Summer 2011 observing season.


\acknowledgements

The Natural Sciences and Engineering Research Council of Canada supports the research of B.C. and R.J.
E.M-R.K.'s work was performed under
contract with the California Institute of Technology funded by NASA
through the Sagan Fellowship Program.
The authors would like to thank the anonymous referee and the great many other
people who commented on and improved this manuscript prior to publication.
The authors especially appreciate the hard-work and diligence of the CFHT staff
for both scheduling these challenging observations and ensuring these ``Staring Mode'' observations were successful.

\end{document}